\documentclass[aip,%
               jcp,%
               preprint,%
               citeautoscript,%
               bibnotes,%
               amsfonts,%
               amssymb,%
               amsmath,%
               a4paper,%
               titlepage,%
               fleqn,%
               endfloats*%
               ]{revtex4-1}

\usepackage{graphicx}
\usepackage{xspace}
\usepackage{float}
\usepackage{color}
\usepackage{subfigure}

\begin{document}

\newcommand{\hoch}[1]{$^{\text{#1}}$}
\newcommand{\tief}[1]{$_{\text{#1}}$}
\def\MP{\text{MP2}\xspace}
\def\gm{\mu}
\def\gn{\nu}
\def\gk{\kappa}
\def\gl{\lambda}
\def\gs{\sigma}
\def\gt{\tau}
\def\ga{\alpha}
\def\gd{\delta}
\def\ogm{\underline{\mu}}
\def\vgm{\bar{\mu}}
\def\vgn{\bar{\nu}}
\def\ogk{\underline{\kappa}}
\def\vgl{\bar{\lambda}}
\def\agk{\tilde{\kappa}}
\def\agl{\tilde{\lambda}}
\def\po{\underline{P}}
\def\pv{\overline{P}}
\def\pa{\widetilde{P}}
\def\ea{\epsilon_a}
\def\eb{\epsilon_b}
\def\ec{\epsilon_c}
\def\ei{\epsilon_i}
\def\ej{\epsilon_j}
\def\ek{\epsilon_k}
\def\et{\epsilon_{\tau}}
\def\1rdm{\gamma}
\def\2rdm{\Gamma}
\def\3rdm{\Gamma}
\def\4rdm{\Gamma}
\def\acr{a^{\dag}}
\def\aan{a}
\def\bra#1{\langle #1|}
\def\ket#1{|#1\rangle}
\def\braket#1#2{\langle #1|#2\rangle}
\def\spA{\left[ 0 \right]}
\def\spB{\left[ -1\right]}
\def\spC{\left[ +1\right]}
\def\spD{\left[ -2\right]}
\def\spE{\left[ +2\right]}
\def\spF{\left[ -1\right]'}
\def\spG{\left[ +1\right]'}
\def\spH{\left[  0\right]'}
\def\nact{N_{\text{act}}}
\def\nele{N_{\text{el}}}

\title{Laplace-transformed multi-reference second-order perturbation theories
       in the atomic and active molecular orbital basis}

\author{Benjamin Helmich-Paris}
\email{b.helmichparis@vu.nl}
\affiliation{Section of Theoretical Chemistry, VU University Amsterdam,
             De Boelelaan 1083, 1081 HV Amsterdam, The Netherlands}

\author{Stefan Knecht}
\affiliation{Laboratory of Physical Chemistry, ETH Z{\"u}rich,
             Vladimir-Prelog-Weg 2, CH-8093 Z{\"u}rich, Switzerland}

\date{\today}

\begin{abstract}
In the present article, we show how
to formulate the partially 
contracted n-electron valence
second order perturbation theory (NEVPT2) energies in the atomic and active
molecular orbital basis by employing the Laplace transformation
of orbital-energy denominators (OED).
As atomic-orbital (AO) basis functions are inherently localized 
and the number of active orbitals is comparatively small,
our formulation is particularly suited for a linearly-scaling NEVPT2 implementation.
Some of the NEVPT2 energy contributions can be formulated completely
in the AO basis as single-reference second-order M{\o}ller--Plesset perturbation
theory and benefit from sparse active-pseudo 
density matrices --- particularly if the active molecular orbitals are localized
only in parts of a molecule.
Furthermore, we show that for multi-reference perturbation theories 
it is particularly challenging to find optimal parameters of the 
numerical Laplace transformation
as the fit range may vary among the 8~different OEDs
by many orders of magnitude.
Selecting the number of quadrature points for each OED
separately according to an accuracy-based criterion allows us
to control the errors in the NEVPT2 energies reliably.
\end{abstract}

\maketitle

\section{Introduction} \label{intro}
Many applications of quantum chemical simulations
require a multi-reference (MR) wave function
to provide an at least qualitatively correct description
of the atomic or molecular system.
Even the simplest molecule --- the hydrogen molecule ---
becomes a MR case in the process
of separating the two hydrogen atoms.
MR theories are also indispensable when two
potential energy surfaces of the same symmetry come
close in energy,
which is frequently encountered when studying the
time evolution of excited electrons in molecules
--- in particular by surface-hopping
excited-state dynamics.\cite{Martinez2006,*Subotnik2016}
Beyond a reliable treatment of bond breaking and potential
energy surfaces, MR theories are often considered
in quantum chemical studies that involve transition metals,
lanthanides, and actinides to cope with open shells
in the electronic structure calculation.
Alternatively, Kohn-Sham density functional theory (DFT)
can often provide meaningful energies and properties
for transition metal complexes with low computational costs
even when standard exchange correlation functionals are
employed.
However, DFT still lacks an ansatz that can improve
systematically on either static or dynamic correlation effects
in a computationally affordable manner.

In comparison to their single-reference (SR) analogues,
wave function-based MR electronic structure methods
are conceptually and computationally much more demanding.
At the multi-configurational (MC) self-consistent field (SCF) level,
the energy needs to be minimized with respect to orbital rotations
and configuration coefficients,
which in many situations requires sophisticated
optimizers with quadratic\cite{Lengsfield1980}
or higher\cite{Werner1980,*Werner1981,*Werner1985} convergence rates.
On the one hand, such algorithms can be incredibly efficient
in terms of number of iterations;
on the other hand, a transformation of the two-electron
integrals from the atomic (AO) to the molecular orbital (MO)
basis is necessary, which scales $\mathcal{O}(N^5)$
with the system size $N$.
Moreover, the MCSCF wave function is expanded in
configuration state functions (CSF)
that can be generated by a complete active space\cite{Roos1980} (CAS).
The latter includes all possible occupations of $\nact$ active MOs
by $\nele$ active electrons
usually denoted by CAS ($\nele$,$\nact$).
With conventional determinant-based full configuration interaction (FCI)
implementations\cite{Knowles1984,Olsen1990} the computational costs
grow exponentially and becomes soon a bottleneck in the MCSCF calculation.
Usually, one is restricted to active spaces
in the ballpark of a CAS (10,30) if no restricted\cite{Olsen1988b} (RAS)
or generalized active spaces\cite{Fleig2001} (GAS) are desired.

There are recent developments in MCSCF
that improve on both computational bottlenecks of the algorithms:
the two-electron integral transformation and
the size limit of the active space.
The latter can be lifted
if the CAS CI secular equations are solved by modern quantum Monte Carlo\cite{Booth2009,Thomas2015}
or density matrix renormalization
group\cite{whit92,*whit93} (DMRG)
algorithms.
This facilitates CASSCF calculations with
much larger active spaces, e.\ g.\ $\nele \times \nact \approx 2000$.\cite{zgid08,*zgid08b,*ghos08,*yana09a,*kurashige2011,*wouters2014,*wout15,*mayi16c,*LiManni2016,knec16a,sunq16b}
As shown lately by Hohenstein et al.\cite{Hohenstein2015}
\ for an approximate second-order optimizer\cite{Chaban1997},
the costly integral transformations can be avoided
by initially computing Coulomb- and exchange-like matrices in the sparse
AO basis with $\mathcal{O}(N^2)$ costs and later transforming to Fock-like intermediates and
orbital gradients in the MO basis.
Just recently, a truly second-order optimizer
that is also AO-driven and can handle a large number of AOs and
active MOs was proposed in Ref.~\citenum{sunq16b}

The computationally simplest approach to account for
dynamic correlation in molecules that demand for
multiple reference determinants
is second-order multi-reference perturbation theory (MRPT2).
MRPT2 methods made substantial progress over the last 25 years
and are frequently used in quantum chemical investigations.
In particular, the CASPT2 method\cite{Andersson1990,*Andersson1992} initially developed by
Andersson et al.\ has evolved into a reliable tool for the investiagation
of molecules with a complicated electronic structure.
Geometries can be conveniently optimized due to availability
of analytic CASPT2 gradients,\cite{Celani2003,*Shiozaki2011,*Gyoerffy2013}
which were developed by Werner and his co-workers.
Accurate CASPT2 energies and gradients near the complete basis set limit
can be obtained with an explicitly correlated F12
variant proposed by Shiozaki et al.\cite{Shiozaki2010,*Shiozaki2013}
For calculations involving heavy elements, spin-orbit coupling (SOC)
can be treated either within a SOC-CI formalism\cite{Roos2004}
or by using the four-component Dirac-Coulomb Hamiltonian.\cite{Abe2006}
Concerning studies of excited states,
the erratic mixing of valence and Rydberg states at the CASSCF level
can be cured by using quasi-degenerate MRPT2.\cite{Finley1998,Angeli2004}

Despite its popularity and success in numerous application,
CASPT2 has two major drawbacks:
(i) It is not size-consistent;
(ii) It is often plagued by intruder states, which complicates
the iterative determination of the first-order wavefunction coefficients.
In practice, the size-inconsistency errors are rather decent
--- even for larger molecules\cite{Menezes2016} --- and the intruder-state problem
can be cured partially by level shifts.\cite{Roos1995,*Forsberg1997}
However, since level shifts alter the second-order energy unpredictably,
CASPT2 should then be considered as a rather empirical method.

The remedies of CASPT2 are cured altogether within a different
formulation of MRPT: the n-electron valence (NEV) PT of Angeli,
Cimiraglia, Malrieu and their co-workers.\cite{Angeli2000,Angeli2001,Angeli2001b,Angeli2002}
NEVPT2 differs from CASPT2 mainly in the definition in the
zeroth-order Hamiltonian.
In NEVPT the Fock operator is replaced by the Dyall Hamiltonian\cite{Dyall1995},
which also accounts for explicit two-electron interactions
within the active orbitals.
Initially, two variants of NEVPT were introduced that differ
in the level of so-called outer-space contraction:\cite{Angeli2001}
the partially contracted (PC) and strongly contracted (SC) NEVPT2.
Both variants are fully internally contracted (IC) and require
the computation of the four-particle reduced density matrix (RDM),
which scales with $\mathcal{O}(\nact^8)$.
With the rise of DMRG and FCIQMC algorithms in the last two decades,
relatively large active spaces are now feasible for CASSCF,
but become prohibitive in the subsequent IC-MRPT calculation.
A rigorous approach to this dilemma would be to use
completely uncontracted first-order wavefunctions,
which was initially considered as computationally not feasible.
Recent developments\cite{Sharma2014,Sokolov2016,Giner2017} reveal,
however, the clear potential of fully uncontracted MRPT
for calculations that demand large active spaces.

Most systems with MR character that have a practical relevance,
e.\ g.\ metal-organic complexes, enzymes, or organic radicals,
have a rather small number of active orbitals
compared to the total number of MOs.
In such situations, the computational costs of MRPT2
are dominated by the $\mathcal{O}(N^5)$ scaling integral transformations
as for SR second-order M{\o}ller--Plesset PT (MP2).
A substantial reduction of the timings can be achieved by employing
a CD of
two-electron integrals\cite{Aquilante2008,*Freitag2017},
in particular, if such an integral decomposition
technique is combined with a truncated or frozen
natural orbital (FNO) virtuals basis.\cite{Aquilante2009}
However, for FNO- or CD-MRPT2 only the pre-factor of the computational
costs is reduced but the scaling is still $\mathcal{O}(N^5)$ as for
conventional implementations.
Thus, calculation that exploit those approximations are still limited to
medium-sized molecules in the ballpark of 100 atoms.
Much larger systems can be calculated if
the first-order wave function is expanded
in a truncated set of pair natural orbitals (PNO),
which was explored recently by the Neese\cite{Guo2016} and
Werner groups\cite{Menezes2016} for PNO-NEVPT2 and PNO-CASPT2, respectively.
Both implementations scale linearly with the system size
and allow to perform MRPT2 calculations of molecules with up to 300 atoms.

In the present article, we pursue an alternative approach to
reach linearly scaling MRPT2 implementations eventually.
It is based on the Laplace transformation of orbital-energy denominators,\cite{Almloef1991}
which was introduced initially by Alml{\"o}f
in the context of MP PT.\cite{Haeser1992,Haeser1993}
In our formulation, all intermediates are given solely in the AO and active MO basis,
which is in line with recent algorithmic developments in MCSCF\cite{Hohenstein2015,sunq16b}.
On the one hand, the number of active MOs $\nact$ can still be considered as small in MR calculations
because even with modern DMRG and FCIQMC implementations solving the CAS CI
secular equations may easily become a computational bottleneck
due to the formal exponential cost scaling with $\nact$.
On the other hand, localized AO basis functions are inherently suited for effective screening.
It was shown by Ayala and Scuseria\cite{Ayala1999} that
SR Laplace-transformed AO-MP2 energies show the physically correct
$R^{-6}$ decay with the distance $R$ between charge distributions.
This was exploited later by Ochsenfeld and his co-workers,\cite{Lambrecht2005a,Doser2009,*Maurer2013b}
who introduced integral estimates with the proper $R$ dependence
to develop linearly scaling MP2 implementations.
With those AO-based implementations MP2 calculations on molecules
with more than 1000 atoms and more than 10000 basis functions were reported.\cite{Doser2009}
However, since the initial AO-MP2 implementation of H{\"a}ser\cite{Haeser1993},
AO-based wave function methods were criticized for their
large computational overhead that increases substantially
for larger basis sets.
This problems can be solved now by a
Cholesky decomposition (CD) of (pseudo-)densities\cite{Zienau2009},
which leads to intermediates that have the
same dimension as the occupied or virtual MO coefficients
but retain the sparsity of AO (pseudo-)density matrices.\cite{Kussmann2015,Luenser2017}
An extension of Laplace-transformed AO-based MP2 for
relativistic all-electron calculations was introduced
recently by one of the present authors for two-component
Hamiltonians and the Kramers-restricted formalism.\cite{Helmich-Paris2016c}

In this work, we show how to formulate NEVPT2 in the
atomic and active molecular orbital basis by means
of the numerical Laplace transformation.
We start in Sec.\ \ref{theory} with a r{\'e}s{\'u}me of present
state-of-the-art IC-MRPT2 theories and argue why
only the partially-contracted NEVPT2 is suitable for
such a re-formulation.
Moreover, we outline how to obtain PC-NEVPT2 energies
in the AO and active MO basis in the spin-free formalism
and provide working equations for each energy contribution.
In Sec.\ \ref{results}, the working equations are discussed
in context of a potentially efficient implementation.
Furthermore, we focus on how to control the error of the
numerical quadrature of the Laplace transformation for
each of the individual energy contributions.
Finally, our procedure is validated for some typical applications
of MRPT methods that involve bond dissociation and excited
valence and Rydberg states.

\section{Theory} \label{theory}

\subsection{R{\'e}s{\'u}me of CASPT2 and NEVPT2}
The zeroth-order wavefunction is given by a linear combination
\begin{align} \label{wpt0}
 \ket{0} \equiv \ket{\Psi^{(0)}} = \sum_K c_K \ket{K}
\end{align}
of all configuration state functions $\ket{K}$.
The wavefunction parameters, that is
orbital rotations and CI coefficients $c_K$, can be
optimized either in a MCSCF or open-shell HF
and subsequent CI calculation.
The first-order correction to the wave function
\begin{align} \label{wpt1a}
& \ket{\Psi^{(1)}} = \sum_{\mu_2} C_{\mu_2} \ket{\Phi_{\mu_2}}
= \sum_{\mu_2} C_{\mu_2} \hat{\tau}_{\mu_2} \ket{0}
\end{align}
is expanded through all possible determinants that are generated
by two-electron excitations from the complete zeroth-order wavefunction.
This approximation is termed internal contraction (IC) and $\ket{\Phi_{\mu_2}}$ 
is referred to as internally contracted configurations\cite{Celani2000} (ICC).
To determine the PT energy through second order,
\begin{align}\label{ept2}
& E^{(2)} = \bra{0} \hat{H} \ket{\Psi^{(1)}}
          = \bra{0} \hat{H} \hat{\tau}_{\mu_2} \ket{0} C^{(1)}_{\mu_2} \text{,}
\end{align}
the expansion coefficients of the first-order wave function $C^{(1)}_{\mu_2}$
are determined by solving the linear system of equations
\begin{align} \label{wpt1}
   \bra{\widetilde{\Phi_{\mu_2}}} \hat{H}^{(0)} - E^{(0)} \ket{\Psi^{(1)}}
 &=  \sum_{\nu_2} \bra{\widetilde{\Phi_{\mu_2}}} [\hat{H}^{(0)},\hat{\tau}_{\nu_2}] \ket{0} C_{\nu_2} \notag \\
 &= -\bra{\widetilde{\Phi_{\mu_2}}} \hat{H} \ket{0}
\text{.}
\end{align}
In Eq.\ \eqref{wpt1} we assume that diagonalization of the zeroth-order Hamiltonian,
\begin{align} \label{ept0}
 & \hat{H}^{(0)} \ket{0} = E^{(0)} \ket{0}
\text{,}
\end{align}
gives the MCSCF or CI eigenvalues $E^{(0)}$ and
zeroth-order wave function $\ket{\Psi^{(0)}}$.
As projection manifold $\{\bra{\widetilde{\Phi_{\mu_2}}}\}$
we choose bra ICCs in Eq.\ \eqref{wpt1}
such that they keep the core and virtual part of the ICC overlaps
$\braket{\widetilde{\Phi_{\mu_2}}}{\Phi_{\nu_2}}$
orthogonal\cite{Pulay1984,*Helgaker2000-biortho}
because simplified arithmetic expressions are obtained.

To build $\ket{\Psi^{(1)}}$, the mutually orthogonal ICCs are
generated by exciting two electrons out of the core
or active into the virtual or active molecular orbitals (MO) of
the zeroth-order wavefunction $\ket{\Psi^{(0)}}$.
There are 8 different classes of ICCs if the reference determinants $\ket{K}$ were built
from a complete active space (CAS), which are compiled in Tab.\ \ref{tab3}.
In the present work, we use spin-free operators that are composed of
singlet one-particle excitation operators
\begin{align}
\hat{E}_{pq} = \sum_{\sigma}\hat{a}^{\dag}_{p\sigma} \hat{a}_{q\sigma}
 \, \text{,}  \quad
 \sigma = \{ \alpha,\beta \}
\end{align}
We follow the convention that core orbitals are denoted by $i,j,k,l$, active orbitals by $t,u,v,w$,
virtual orbitals by $a,b,c,d$, and general orbitals by $p,q,r,s$.
The different classes of ICCs in Tab.\ \ref{tab3} are labeled by symbols
that became standard notation in NEVPT:
$+n$ denotes attachment of $n$ electron(s)
into the active space while $-n$ means ionization of $n$ electron(s)
from the active space;
a prime designates a single-particle excitation
within the active orbital space.
In contrast to other ICCs, the one of the $\spH$ class are composed
of two kinds of excited determinants,  $\ket{\Phi_{it}^{au}}$  and $\ket{\Phi_{it}^{ua}}$,
that are mutually non-orthogonal.
If RAS or GAS were employed in the underlying MCSCF calculation,
the expansion of $\ket{\Psi^{(1)}}$ in terms of ICCs
needs to be augmented with a subset of the CAS space determinants, i.\ e.
$\ket{\Phi_{tv}^{uw}} = \hat{E}_{ut} \hat{E}_{wv} \ket{0}$,
to keep the zeroth- $\ket{\Psi^{(0)}}$ and first-order wavefunction
$\ket{\Psi^{(1)}}$ mutually orthogonal.\cite{Celani2000}
In the present work, we only consider CAS reference wavefunctions.

In MRPT the definition of the zeroth-order Hamiltonian $\hat{H}^{(0)}$ is not unique
and it is the choice of $\hat{H}^{(0)}$ in which CASPT2 and PC-NEVPT2 mainly differ.
In CASPT2 the zeroth-order Hamiltonian
\begin{align} \label{h0cas}
  \hat{H}^{(0)} &= \hat{P} \hat{F} \hat{P} + \hat{Q} \hat{F} \hat{Q}
\end{align}
is defined by projection of the Fock operator
\begin{align}
 \hat{F} &= \sum_{pq} f_{pq} \hat{E}_{pq} \text{,} \\
 f_{pq} &= f^{\text{I}}_{pq}+{^A}f_{pq}  \text{,} \label{focktot} \\
 f^{\text{I}}_{pq} &= h_{pq} + \sum_i \left[ 2 (pq|ii) - (pi|iq) \right] \text{,} \\
 f^{\text{A}}_{pq} &= \frac{1}{2} \sum_{tu} \left[ 2 (pq|tu) - (pu|tq)  \right] \1rdm_{tu} \text{,}
\end{align}
onto the space of reference determinants
$ \hat{P} = \ket{0} \bra{0}$ and its complement $\hat{Q} = \hat{1} - \hat{P}$.\cite{Celani2000}
The inactive $f^{\text{I}}_{pq}$ and active $f^{\text{A}}_{pq}$ Fock matrix\cite{Helgaker2000-fock} are built
from the one- $h_{pq}$ and two-electron integrals $g_{pqrs}$, as they appear in the electronic Hamiltonian
\begin{align}
 \hat{H} &= \sum_{pq} h_{pq} \hat{E}_{pq} +
  \frac{1}{2} \sum_{pqrs} g_{pqrs} (\hat{E}_{pq} \hat{E}_{rs} - \gd_{qr} \hat{E}_{ps} ) \text{,} \\
 h_{pq}  &= \sum_{\gm\gn}       C_{\mu p} C_{\nu q} h_{\gm\gn}  \text{,} \\
 g_{pqrs} &\equiv (pq|rs) = \sum_{\gm\gn\gk\gl} C_{\gm p} C_{\gn q} C_{\gk r} C_{\gl s} (\gm\gn|\gk\gl)
\text{,}
\end{align}
and the singlet one-particle reduced density matrix (1-RDM) $\1rdm_{tu}$ (see Tab.\ \ref{tab2}).
The integrals in the AO basis $h_{\gm\gn}$ and $(\gm\gn|\gk\gl)$ are
transformed to the corresponding MO integrals $h_{pq}$ and $g_{pqrs}$ by
MO coefficients $C_{\gm p}$ that are available from the preceding
MCSCF or open-shell HF calculation.
The CI coefficients appear only in the n-RDMs, as given in Tab.\ \ref{tab2},
for IC PTs, which are our focus here.

Due to the non-diagonal block structure of $\hat{Q} \hat{f} \hat{Q}$ in CASPT2,
$\hat{H}^{(0)}$ can couple different classes of ICCs
in $\bra{\widetilde{\Phi_{\mu_2}}} [\hat{H}^{(0)}, \hat{\tau}_{\nu_2}] \ket{0}$.
This has two unpleasant implications:
first, Eq.\ \eqref{wpt1} must be determined iteratively
and a larger computational overhead compared to a direct method like single-reference MP2
is expected;
second, CASPT2 may not be size extensive\cite{VanDam1998,*VanDam1999}
what, in principle, disqualifies the method to be applied to large molecular systems.
Furthermore, the usage of the Fock operator not only for core and virtual
but also for the active orbitals may lead to (near) singularities
in Eq.\ \eqref{wpt1}.
In particular, ICC blocks in
$\bra{\widetilde{\Phi_{\mu_2}}} [\hat{H}^{(0)}, \hat{\tau}_{\nu_2}] \ket{0}$
that involve three active orbitals, $[-1]'$ and  $[+1]'$,
are most likely to become zero.\cite{Dyall1995,Angeli2000}
This is the root of the notorious CASPT2 intruder state problem,
which can be cured either by including more orbitals
into the active space or in an empirical fashion by introducing level shifts
in the zeroth-order Hamiltonian\cite{Roos1995,*Forsberg1997}.
Furthermore, near singularities in $\bra{\widetilde{\Phi_{\mu}}} [\hat{H}^{(0)}, \hat{\tau}_{\nu}] \ket{0}$
would impede its Laplace transform as many quadrature points 
in the numerical integration procedure would be required then.

The NEVPT2 zeroth-order Hamiltonian features only diagonal ICC blocks,\cite{Dyall1995}
\begin{align} \label{h0nev}
 \hat{H}^{(0)} &= \hat{P} \hat{H}^{\text{D}} \hat{P}  +
 \hat{P}^{\spA} \hat{H}^{\text{D}} \hat{P}^{\spA}  + \ldots +
 \hat{P}^{\spH} \hat{H}^{\text{D}} \hat{P}^{\spH}
\text{,}
\end{align}
which can guarantee size extensivity.\cite{VanDam1998,*VanDam1999}
Furthermore, Dyall's Hamiltonian $\hat{H}^{\text{D}}$ is employed in Eq.\ \eqref{h0nev} that
chooses the core $\hat{H}_{\text{c}}$ and valence part $\hat{H}_{\text{v}}$ differently,
\begin{align}
 \hat{H}^{\text{D}} &= \hat{H}_{\text{c}} + \hat{H}_{\text{v}} + C  \text{,} \\
 \hat{H}_{\text{c}} &= \sum_{ij} f_{ij} \hat{E}_{ij} + \sum_{ab} f_{ab} \hat{E}_{ab}  \text{,} \\
 \hat{H}_{\text{v}} &= \sum_{tu} f^{\text{I}}_{tu} \hat{E}_{tu} +
   \frac{1}{2} \sum_{tuvw} (tu|vw) (\hat{E}_{tu} \hat{E}_{vw} - \gd_{uv} \hat{E}_{tw} )
\text{,}
\end{align}
and accounts for two-electron
interactions among the active electrons.\cite{Dyall1995,Angeli2000,Angeli2002}
The constant $C$ is chosen such that Eq.\ \eqref{ept0} is fulfilled.

The bi-electronic valence part of Dyall's Hamiltonian $\hat{H}_{\text{v}}$
in the commutator of Eq.\ \eqref{wpt1}
$\bra{\widetilde{\Phi_{\gm_2}}}  [\hat{H}_{\text{v}},\hat{\tau}_{\nu_2}] \ket{0}$
leads to the so-called Koopmans matrices.
For example, the $\spB$ Koopmans matrix is given as
\begin{align}
K^{\spB}_{t't} &=
\frac{1}{2} \bra{0} E_{t'b} E_{ia} [ \hat{H}_{\text{v}}, E_{ai} E_{bt} ] \ket{0} \\
&= - \sum_{tu} f^{\text{I}}_{tu} \1rdm_{tu} - \sum_{uvw} (tu|vw) \2rdm_{t'v,uw}
 \label{koop-1}
\end{align}
and represents a single-ionization potential.\cite{Morrell1975,Smith1975,Angeli2001}
In Eq.\ \eqref{koop-1} $\2rdm_{t'v,uw}$ is the singlet 2-RDM,
which is defined in Tab.\ \ref{tab2}.
The term Koopmans matrices for classes denoted with a prime
is chosen for reasons of notational
convenience rather than physical rationality.
Explicit expressions for all 7 kinds of Koopmans matrices
are presented in the spin-free formalism in Ref.\ \onlinecite{Angeli2002} and not repeated here.
For large active spaces the computation of the $\spF$  and $\spG$
Koopmans matrices can be demanding since it requires prior
calculation of the 4-RDM (see Tab.\ \ref{tab2}).

The iterative solution of the linear system of equations \eqref{wpt1}
can be avoided in the partially contracted (PC) NEVPT2 variant
by solving a generalized eigenvalue problem (GEP),
\begin{align} \label{koopgep}
& \mathbf{K}\, \mathbf{c} = \mathbf{M}\, \mathbf{c}\, \boldsymbol{\epsilon}
\text{,}
\end{align}
for each excitation class separately.
Eq.\ \eqref{koopgep} involves the Koopmans matrices $\mathbf{K}$
and the active part of the ICC overlap
$\braket{\widetilde{\Phi_{\mu_2}}}{\Phi_{\nu_2}}$,
which is denoted here as metric matrix.
The metric matrices $\mathbf{M}$ for all excitation classes
are compiled in Tab.\ \ref{tab2}
and formulated in terms of singlet RDMs.
If a CAS zeroth-order wave function is employed,
both $\mathbf{K}$ and $\mathbf{M}$
are Hermitian and, thus, the eigenvalues $\boldsymbol{\epsilon}$ are necessarily real
for each of the 7 classes.\cite{Angeli2002}
Furthermore, the Koopmans matrices $\mathbf{K}$ are either positive
definite for those classes representing active space electron attachment or excitation
or negative definite for the active electron ionization classes.
Consequently, in the first-order equation \eqref{wpt1},
$\bra{\widetilde{\Phi_{\mu_2}}} [\hat{H}^{(0)}, \hat{\tau}_{\nu_2}] \ket{0}$
can never become singular, that is, there are no intruder states in NEVPT2.
This facilitates a direct calculation of the PC-NEVPT2 energy with
a formulation that is suited for a Laplace transformation of denominators
involving Fock and Koopmans matrix eigenvalues,
as we elaborate on in Sec.\ \ref{reform}.

Apart from choosing a IC first-order wave function in Eq.\ \eqref{wpt1a}, as done
in CASPT2 and PC-NEVPT2, a further level of contraction can be introduced
by choosing an alternative excitation operator basis for the ICCs.
For the strongly contracted (SC) NEVPT2, the active orbitals of the two particle excitations operators
are contracted with parts of the perturbation operator\cite{Angeli2001} as shown in Tab.\ \ref{tab3}
for the bra ICCs.
This leads to intermediates $\boldsymbol{\epsilon}$ in denominators of the energy expressions
that have the same dimension as the SC ICCs,\ e.\ g.\ one obtains
an effective quasi-hole energy
\begin{align} \label{effspb}
 & \epsilon_{ab,i}^{\spB} =
 \frac{\bra{\widetilde{\Phi_i^{ab}}} \sum_t [\hat{H}_{\text{v}},\hat{E}_{ai} \hat{E}_{bt}] \ket{0}  g_{aibt} }
      {\braket{\widetilde{\Phi_i^{ab}}}{\Phi_i^{ab}}}
\end{align}
for the $\spB$ class.
Such effective energies, as in Eq.\ \eqref{effspb}, are highly unsuited
for a Laplace transformation of the denominators
since, unlike the MP denominators, they cannot be partitioned into individual orbital contributions.
Thus, it is not worth to pursue SC-NEVPT2 further in the present work.

\subsection{Re-formulation of PC-NEVPT2 in the atomic and active orbital basis}
\label{reform}
In the following we will present a re-formulation of PC-NEVPT2 energies
in the atomic and active molecular orbital basis.
Only a few working equations of PC-NEVPT2 in the MO basis are presented
to discuss the re-formulation in sufficient detail.
A complete presentation of the conventional
MO-based NEVPT2 can be found in Ref.\ \onlinecite{Angeli2002}.

The $\spA$ energy term is identical to the MP2 energy that depends only on the core and
virtual orbitals:
\begin{align} \label{espA}
 E^{[0]} &= - \sum_{aibj} \frac{[ 2 (ai|bj) - (aj|bi) ] (ia|jb)}{\ea - \ei + \eb - \ej}
 \text{.}
\end{align}
It can be expressed in terms of orbital-energy denominators (OED) if a canonical MO basis
is employed, which keeps the the core-core and virtual-virtual block
of the Fock matrix in Eq.\ \eqref{focktot} diagonal,
\begin{align}
 \sum_k f_{ik} \, U_{kj} &= U_{ij} \, \ej \text{,} \label{focko} \\
 \sum_c f_{ac} \, V_{cb} &= V_{ab} \, \eb          \label{fockv}
 \text{.}
\end{align}
Such kind of canonical core and virtual MOs are used in the remainder of this article.

The Laplace transformation of OEDs $\Delta = \ea - \ei + \eb - \ej$,
\begin{align} \label{laplace}
\frac{1}{\Delta} &= \int_0^{\infty} \exp(-\Delta\, t) dt
 \approx \sum_{\ga=1}^{n_{\ga}} \omega_{\ga}\, \exp(-\Delta\, t_{\ga})
\text{,}
\end{align}
is employed to factorize the $\spA$ energy term.\cite{Almloef1991,Haeser1992}
The parameters of the numerical quadrature $\{ \omega_{\ga}, t_{\ga} \}$ in \eqref{laplace}
are usually obtained either by least-square minimization\cite{Haeser1992}
or by using the minimax approximation.\cite{Takatsuka2008,Helmich-Paris2016d}
Eq.\ \eqref{laplace} is the foundation of a re-formulation of $\spA$ (or MP2) energies
\begin{align}
E^{\spA} &= -  \sum_{\ga}\, \sum_{\gm\gn\gk\gl}
 (\ogm \vgn|\gk\gl)^{(\ga)} \Big[ 2 (\gm\gn|\ogk\vgl)^{(\ga)} - (\gm\vgl|\ogk\gn)^{(\ga)} \Big]
\end{align}
in terms of intermediates in the AO basis that are obtained
from two consecutive one-index transformation of the two-electron integrals
\begin{align}
(\ogm \vgn|\gk\gl)^{(\ga)}
&= \sum_{\gm'} \po_{\gm'\gm}^{(\ga)} \sum_{\gn'} \pv_{\gn\gn'}^{(\ga)}\, (\gm'\gn'|\gk\gl) \text{,} \\[0.5em]
(\gm \vgl|\ogk\gn)^{(\ga)}
&= \sum_{\gk'} \po_{\gk'\gk}^{(\ga)} \sum_{\gl'} \pv_{\gl\gl'}^{(\ga)}\, (\gm\gl'|\gk'\gn)
\end{align}
with the core and virtual pseudo-density matrices
\begin{align}
 \po_{\gm\gm'}^{(\ga)}
&= |\omega_{\ga}|^{1/4}\, \sum_{i} C_{\gm i}\, \sum_{k} U_{ik}\,  e^{+\ek t_{\ga}}\,\sum_{i'}  U_{i'k}\,  C_{\gm'i'} \text{,} \label{psdeno} \\
 \pv_{\gm\gm'}^{(\ga)}
&= |\omega_{\ga}|^{1/4}\, \sum_{a} C_{\gm a}\, \sum_{c} V_{ac}\,  e^{-\ec t_{\ga}}\,\sum_{a'} V_{a'c}\,  C_{\gm'a'} \label{psdenv}
\text{.}
\end{align}

The energy and wavefunction coefficients of the $\spB$ contribution
in the MO basis is given by
\begin{align}
 &E^{\spB} = \sum_{aibt} C_{it}^{ab} \sum_{u} \1rdm_{tu} [ 2 (ia|ub) - (ib|ua) ] \text{,} \\
 &\sum_u \left[ (\ea - \ei + \eb) \1rdm_{tu} - K^{\spB}_{ut} \right] C_{iu}^{ab}
  = - \sum_u (ai|bu) \1rdm_{tu} \label{wfB}
\end{align}
Solving the linear system of equations \eqref{wfB} is avoided
if the  $\spB$ Koopmans and metric matrix are diagonalized in the GEP
\begin{align}
& \sum_{t'} K^{\spB}_{tt'} c^{\spB}_{t'\tau} = \sum_{t'} \1rdm_{tt'} c^{\spB}_{t'\tau} \et^{\spB}  \label{evpB}
\text{.}
\end{align}
This leads to a direct expression for the $\spB$ energy contribution
\begin{align} \label{espBmo}
&E^{\spB} = - \sum_{aib\tau} \frac{(ai|b\tau)[ 2 (ia|\tau b) - (ib|\tau a) ]}
  {\ea - \ei + \eb - \et^{\spB}} \text{,} \\
& (ai|b\tau) = \sum_u (ai|bu) \sum_{u'} \1rdm_{u'u} c^{\spB}_{u'\tau} \text{,}
\end{align}
with an OED that contains $\spB$ Koopmans matrix eigenvalues.
The number of eigenvalues $\tau$ in the equations above
may be smaller than the number of active orbitals if
$\boldsymbol{\1rdm}$ becomes singular and some pairs of
singular values and vectors of $\boldsymbol{\1rdm}$
need to be removed for reasons of numerical stability (\textit{vide infra}).
To obtain an energy expressions solely in the AO and active MO basis,
like for the $\spA$ (or MP2) term, core and virtual MO coefficients
and Fock matrix eigenvalues in Eqs.\ \eqref{focko} and \eqref{fockv}
can be incorporated into core (Eq.\ \eqref{psdeno}) and
virtual (Eq.\ \eqref{psdenv})  pseudo-density matrices.
Concerning the active-orbital part in Eq.\ \eqref{espBmo},
it is convenient to summarize all those quantities
that appear in GEP \eqref{evpB}
into a single intermediate
\begin{align}
\mathcal{K}^{(\ga),\spB}_{tu}
&=  |\omega_{\ga}|^{1/4} \sum_{t'} \1rdm_{tt'} \sum_{\gt} c^{\spB}_{t',\gt}\,
   e^{+\et^{\spB} t_{\ga}}\sum_{u'} c^{\spB}_{u',\gt}\, \1rdm_{uu'}
\text{,}
\end{align}
which we refer to as Koopmans matrix pseudo-exponential.
By means of $\mathcal{K}^{(\ga),\spB}_{tu}$, the $\spB$ energy
can be formulated in terms of intermediates that depend only
on the AOs and active MOs:
\begin{align}
E^{\spB}
&= -  \sum_{\ga} \sum_{\substack{\gm\gn\gl \\ uu'}} \,
(\ogm\vgn|u\gl)^{(\ga)} \mathcal{K}^{(\ga),\spB}_{uu'} \Big[ 2 (\gm\gn|u'\vgl)^{(\ga)} - (\gm\vgl|u'\gn)^{(\ga)} \Big]
\text{.}
\end{align}
Alternatively, a pure AO-based formulation,
which is more in-line with standard AO-MP2 formulations,
\begin{align}
&E^{\spB} = - \sum_{\ga}\,
 (\ogm\vgn|\gk\gl)^{(\ga)} \Big[ 2 (\gm \gn|\agk^{\spB} \vgl)^{(\ga)} - (\gm\vgl|\agk^{\spB}\gn)^{(\ga)} \Big]
 \text{,} \\[0.5em]
&(\gm \gn|\agk^{\spB}\vgl)^{(\ga)}
  = \sum_{\gk'} \pa_{\gk\gk'}^{(\ga),\spB} \sum_{\gl'} \pv_{\gl\gl'}^{(\ga)}\, \, (\gm\gn|\gk'\gl')
\text{,}
\end{align}
can be chosen by incorporating the Koopmans matrix pseudo-exponential
into an active pseudo-density matrix
\begin{align}
 \pa_{\gm\gm'}^{(\ga),\spB}
&= \sum_u C_{\gm u} \sum_{u'} \mathcal{K}^{(\ga),\spB}_{uu'}\, C_{\gm'u'}
\text{.}
\end{align}
Likewise, the energy term $E^{\spC}$ (Tab.\ \ref{tab1})
can be formulated completely in the AO basis by replacing
the virtual pseudo-density matrix
in one of the half-transformed integrals
with an active pseudo-density matrix $\pa_{\gm\gm'}^{(\ga),\spC}$ (Tab.\ \ref{tab1})
that includes the Koopmans matrix eigenvalues describing
single-electron attachment into the active space.

The remaining $E^{(2)}$ energy contributions are compiled in Tab.\ \ref{tab1}
along with their intermediates.
For each class, Koopmans matrix pseudo-exponentials of the form
\begin{align} \label{koopexp}
& \boldsymbol{\mathcal{K}}^{\ga} =
|\omega_{\ga}|^{n/4}\,  \mathbf{M}\, \mathbf{c}\, e^{-\beta\,t_{\ga}\, \boldsymbol{\epsilon} }\,
 \mathbf{c}^T\, \mathbf{M}^T \\[0.5em]
& \beta = \left\{
    \begin{array}{rl}
      +1 & \text{if attachment and/or excitation} \\
      -1 & \text{if ionization}
    \end{array}
 \right.
\end{align}
are computed.
In Eq.\ \eqref{koopexp} $n$ is the number of active orbital pairs in metric $\mathbf{M}$
and $\beta$ the sign with that the Koopmans matrix eigenvalues
$\boldsymbol{\epsilon}$ enter the OEDs.
In analogy to AO-MP2, where the core and virtual
pseudo-density matrices can be computed without a
preceding diagonalization of the Fock matrix in the SCF procedure,\cite{Surjan2005}
one could consider a formulation of the Koopmans matrix pseudo-exponential
where solving the GEP \eqref{koopgep} is avoided,
\begin{align} \label{koopexp2}
\boldsymbol{\mathcal{K}}^{\ga} &=
 |\omega_{\ga}|^{n/4} \, \mathbf{M} e^{-\beta\,t_{\ga}\, \mathbf{M}^{-1}\, \mathbf{K}}
\text{.}
\end{align}
However, in case of the pseudo-density matrices in AO-MP2, the
inverse of the metric matrix, that is the AO overlap $\mathbf{S}$,
is readily available in form of the core $\mathbf{P}$ and
virtual density matrix $\mathbf{Q}$:\cite{Ayala1999,Surjan2005}
\begin{align}
& \mathbf{S}^{-1} = \mathbf{P} + \mathbf{Q}
\text{.}
\end{align}
In Eq.\ \eqref{koopexp2} the inverted metric matrix is initially
unknown and must be determined by diagonalization
and subsequent removal of linearly dependent eigenvectors.
Consequently, it seems that there is no gain in computational performance if
one tries to circumvent solving the GEP \eqref{koopgep}.

The explicit expressions for the second-order energy contributions in Tab.\ \ref{tab2}
show a rather surprising resemblance with the recently proposed
time-dependent (t) NEVPT2 of Sokolov and Chan.\cite{Sokolov2016}
Apart from an AO-based formulation for the core and virtual orbitals,
the two formulations merely differ in the active orbital-based
intermediates.
To obtain explicit expressions for our Laplace-transformed PC-NEVPT2 formulation
from those of t-NEVPT2,
the time-ordered m-particle 1-time Green's functions with (m=1-3),\cite{Sokolov2016}
e.\ g.\ for the $\spD$ space
\begin{align}
 G_{tv,uw}(t) &= \sum_{\gs\gs'} \bra{0} \acr_{t\gs}(t)\, \acr_{v\gs'}(t)\, \aan_{w\gs'}\, \aan_{u\gs} \ket{0}
\text{,}
\end{align}
needs to be replaced by the corresponding time-dependent
Koopmans matrix pseudo exponentials
\begin{align} \label{koopexp-2}
& \boldsymbol{\mathcal{K}}(t) =
 \mathbf{M}\, \mathbf{c}\, e^{-\beta\,t\, \boldsymbol{\epsilon} }\,
 \mathbf{c}^T\, \mathbf{M}^T
\text{;}
\end{align}
if we consider the analytic form of the Laplace transformation in Eq.\ \eqref{laplace}.

\section{Computational details}\label{comp}
We used the CASSCF procedure as implemented in the MOLCAS package\cite{MOLCAS8.0} to obtain 
the zeroth-order MOs. 
Subsequently, Koopmans matrices were generated by the \textsc{relmrpt2} module, a locally modified version of a
DMRG-NEVPT2 module \cite{Freitag2017}, from n-RDMs\ obtained by means of DMRG calculations with the 
\textsc{QCMaquis} package\cite{Keller2015,kell16,knec16a}. 

A pilot-implementation of the Laplace-transformed PC-NEVPT2
in the atomic and active molecular orbital basis was
integrated into the Kramers-restricted two-component AO-MP2\cite{Helmich-Paris2016c}
implementation,
which is part of a development version of the DIRAC package.\cite{DIRAC16}
The parameters of the numerical quadrature were obtained
by using an implementation of the minimax approximation\cite{Takatsuka2008,Helmich-Paris2016d},
which is provided as external open-source library\cite{laplace-minimax}.
The two-electron integrals needed for NEVPT2 were calculated with
the InteRest library.\cite{InteRest2.0}

Except for the $\spB$ and $\spC$ class, the GEP \eqref{koopgep}
is plagued by numerically instabilities caused by singularities in the
metric matrix $\mathbf{M}$.
To guarantee numerical stability when solving the GEP \eqref{koopgep},
the eigenvalues of $\mathbf{M}$ are computed first
and then compared to a threshold to remove singular value-vector pairs
by a canonical orthogonalization procedure.\cite{Szabo1996}
The linear dependency threshold is set to $10^{-6}$ for all classes
as in the implementation described in Ref.\ \onlinecite{Angeli2001b,Aidas2014}.
Point-group (PG) symmetry cannot be exploited for our NEVPT2 implementation
at the current stage.
Consequently, C\tief{1} PG symmetry was enforced for all CASSCF
and DMRG calculations, though all molecules investigated in the 
this work have at least two symmetry elements.

All results presented here were obtained by employing the full Fock
matrix in the definition of the core part of the Dyall Hamiltonian\cite{Angeli2004b,Havenith2004}
rather than only the diagonal elements\cite{Angeli2001}
to guarantee rotational invariance within all three orbital spaces.
However, we omitted the off-diagonal elements in the Fock matrix
when validating the correctness of the working equations in Tab.\ \ref{tab1}
and our implementation by comparing our results with the ones
from the NEVPT2 implementation in DALTON.\cite{Angeli2001b,Aidas2014}

\section{Results and discussion} \label{results}

\subsection{Discussion of the working equations}
In the previous section, we outlined the derivation of the 
Laplace-transformed PC-NEVPT2 equations
formulated in the AO and active MO basis and presented
explicit expressions in Tab.\ \ref{tab2}.
We proposed that the energy contributions that depend on
integrals with either no or a single active MO, that is
the $\spA$, $\spB$, and $\spC$ term,
can be formulated completely in the AO basis similar to
the SR AO-based MP2\cite{Haeser1993}.
Those three energy terms differ merely in the pseudo-density matrices;
the MP2-like $\spA$ term requires only the core and virtual pseudo-density matrices,
one active pseudo-density matrix
replaces one core and one virtual 
for the $\spB$ and $\spC$ energy term, respectively.

A Laplace-transformed AO-driven implementation of 
the $\spA$, $\spB$, and $\spC$ $E^{(2)}$ terms
increases the computational costs noticeably because,
first, the number of basis functions is much larger than the
number of core and active orbitals and, second,
the time-determining steps have to be repeated
for every quadrature point of Laplace transformed OEDs.
However, all intermediates are formulated in the AO basis
and become sparse for sufficiently large systems.
Furthermore, it was shown for SR AO-MP2
that linearly scaling implementations with an early onset\cite{Doser2009,*Maurer2013b}
can be obtained as the contributions of the AO quadruple
to the direct or Coulomb MP2 energy decay as $R^{-4}$
with respect to the separation of electronic charge distributions $R$.\cite{Ayala1999} 
This rapid decay could be attributed to
the vanishing overlap of core and virtual pseudo-density matrices,
\begin{align}
 & \boldsymbol{\po}^{\alpha}\, \boldsymbol{S}\, \boldsymbol{\pv}^{\alpha} = \boldsymbol{0}
\text{,}
\end{align}
in the multipole expansion of the integrals in orders of $R$
This leads to a leading $R^{-6}$ dipole-dipole integral term for the direct MP2 energy.
Since the overlap of core and active respectively virtual and active
pseudo-density matrices vanishes as well,
\begin{align}
 \begin{array}{ccc}
  \boldsymbol{\pv}^{\alpha}\, \boldsymbol{S}\, \boldsymbol{\pa}^{\alpha,\spB} = \boldsymbol{0} &
 \text{ and }  &
  \boldsymbol{\po}^{\alpha}\, \boldsymbol{S}\, \boldsymbol{\pa}^{\alpha,\spC} = \boldsymbol{0}
 \end{array}
\text{,}
\end{align}
due to the mutual orthogonality of the three separate orbital spaces,
the $\spB$ and $\spC$ terms reveal the same $R^{-6}$ decay
of the direct energy contributions as AO-MP2.
This beneficial long-range behavior of the $E^{(2)}$
contributions has been exploited already by Guo\ et\ al.\ 
in their linearly scaling PNO-NEVPT2 implementation 
at the level of electron pair pre-screening.\cite{Guo2016,Pinski2015}

Though the $\spA$, $\spB$ and $\spC$ energy terms show the
same inter-electronic decay behavior, their magnitude
may differ substantially and is sensitive to the number of electrons and
orbitals in the active space.
As $E^{\spB}$ and $E^{\spC}$ depend linearly on
their respective active pseudo-density matrix,
we show contour plots of $\boldsymbol{\pa}^{\spB}$ and $\boldsymbol{\pa}^{\spC}$
along with $\boldsymbol{\po}$ and $\boldsymbol{\pv}$
for the polyene C\tief{32}H\tief{34} (PE-32) and polyene glycol biradical
$\dot{\text{O}}-\text{C}_{32}\text{H}_{64}-\dot{\text{O}}$ (PEG-32)
in Figs.\ \ref{fig2} and \ref{fig3}, respectively.\cite{si-struct}
The calculations were performed with the SVP basis set\cite{Schaefer1992}
and the $1s$ orbitals of the C and O atoms were kept frozen.
For PE-32 the CAS is composed of 8~electrons that are
distributed amongst 8~active valence MOs.
The latter are delocalized over the entire conjugated pi-system of the polyene.
As can be see from Fig.\ \ref{fig2}, 
the active pseudo-density matrices 
$\boldsymbol{\pa}^{\spB}$ and $\boldsymbol{\pa}^{\spC}$
of the polyene are very similar and have their maxima along the diagonal
of the C-C blocks.
The elements of the H-H and H-C blocks of are much smaller in $\boldsymbol{\pa}$.
All $s$-type basis function do not contribute
to the pi-orbitals in the CAS for symmetry reasons and, therefore,
$\boldsymbol{\pa}$ is much sparser than $\boldsymbol{\po}$ and $\boldsymbol{\pv}$.
For the PEG-32 biradical we build a CAS from the 10~electrons
and 6~$2p$ orbitals of the two O atoms.
Those active orbitals are localized at the ends of PEG-32 and
have natural occupation numbers of $n_{\text{NO}}=(2.00,2.00,2.00,2.00,1.00,1.00)$.
As can be seen from Fig.\ \ref{fig3},
the core and virtual pseudo-density matrices $\boldsymbol{\po}$ and $\boldsymbol{\pv}$
have their largest elements at the diagonal shell pairs and their nearest neighbors
whereas the largest elements of the active pseudo-densities $\boldsymbol{\pa}$ 
are located at the O atoms.
As expected from $n_{\text{NO}}$,
$\boldsymbol{\pa}^{\spB}$ has much more significant elements
than $\boldsymbol{\pa}^{\spC}$.
Concerning $\boldsymbol{\pa}^{\spC}$, only those basis functions are important that
have sizable coefficients for the two active MOs with $n_{\text{NO}}=1.00$.

The test molecules that we investigated have different
physical characteristics of their valence orbitals in the active space.
While for polyene the valence MOs are delocalized over nearly the
whole molecule, for the polyethylene glycol biradical the
valence MO are localized at the ends of the linear molecule.
For both systems, we observed that the number of significant elements
in the active pseudo-density matrix is much smaller than in
the core and virtual pseudo-density matrix.
In combination with the discussed $R^{-6}$ decay, 
this should pave the way for a reduced or even linearly scaling implementation
where the ${\spB}$ and ${\spC}$ energies are formulated completely in the AO-basis.
The computational overhead of an entirely AO-based formulation
of the ${\spB}$ and ${\spC}$ energies can be reduced drastically
when performing a CD\cite{Zienau2009,Kussmann2015,Luenser2017} 
of the active pseudo-density matrices.
Such a CDD-based algorithm would combine successfully the sparsity
with the low rank of active pseudo-density matrices.

\subsection{Accuracy of the numerical quadrature}
\begin{figure}[p!]
 \centering
 \subfigure[]{
  \centering
  \label{fig1a}
  \includegraphics[width=0.45\textwidth]{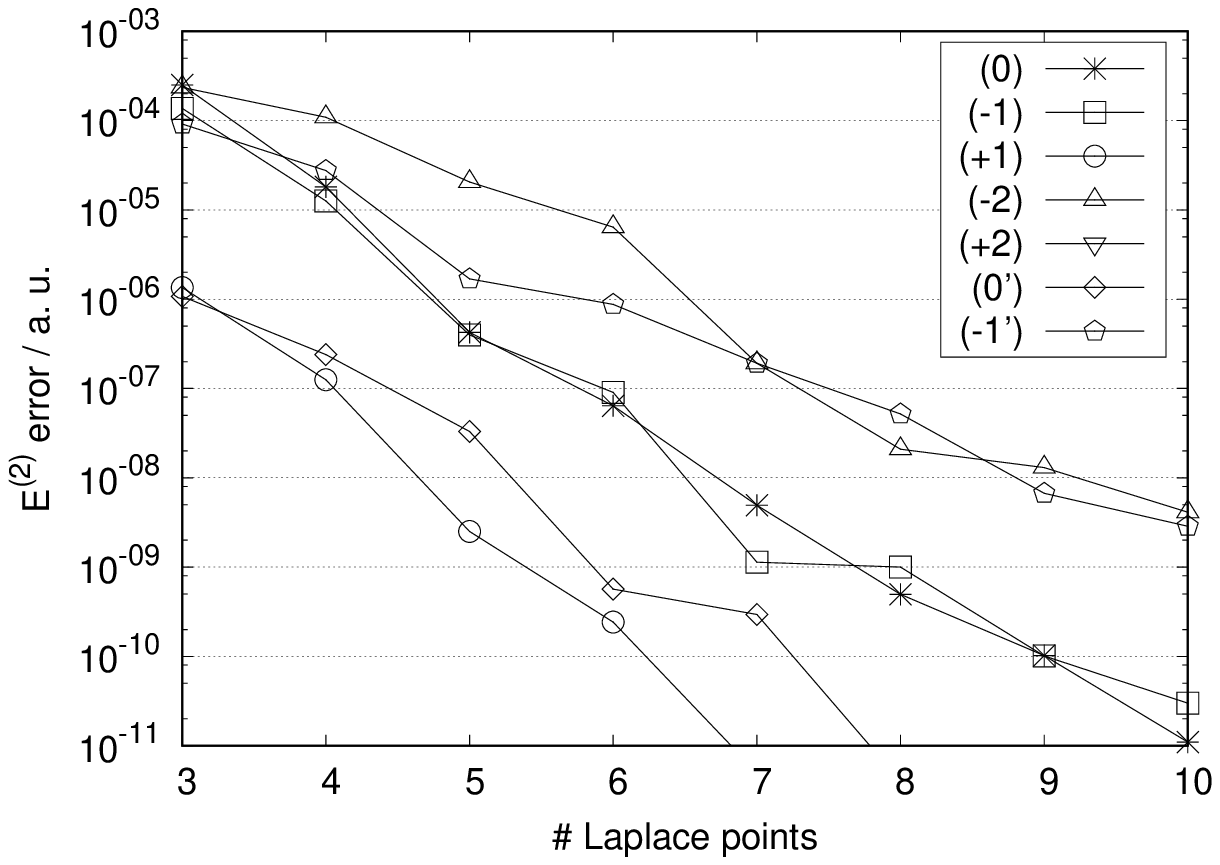}
 }
 \subfigure[]{
  \centering
  \label{fig1b}
  \includegraphics[width=0.45\textwidth]{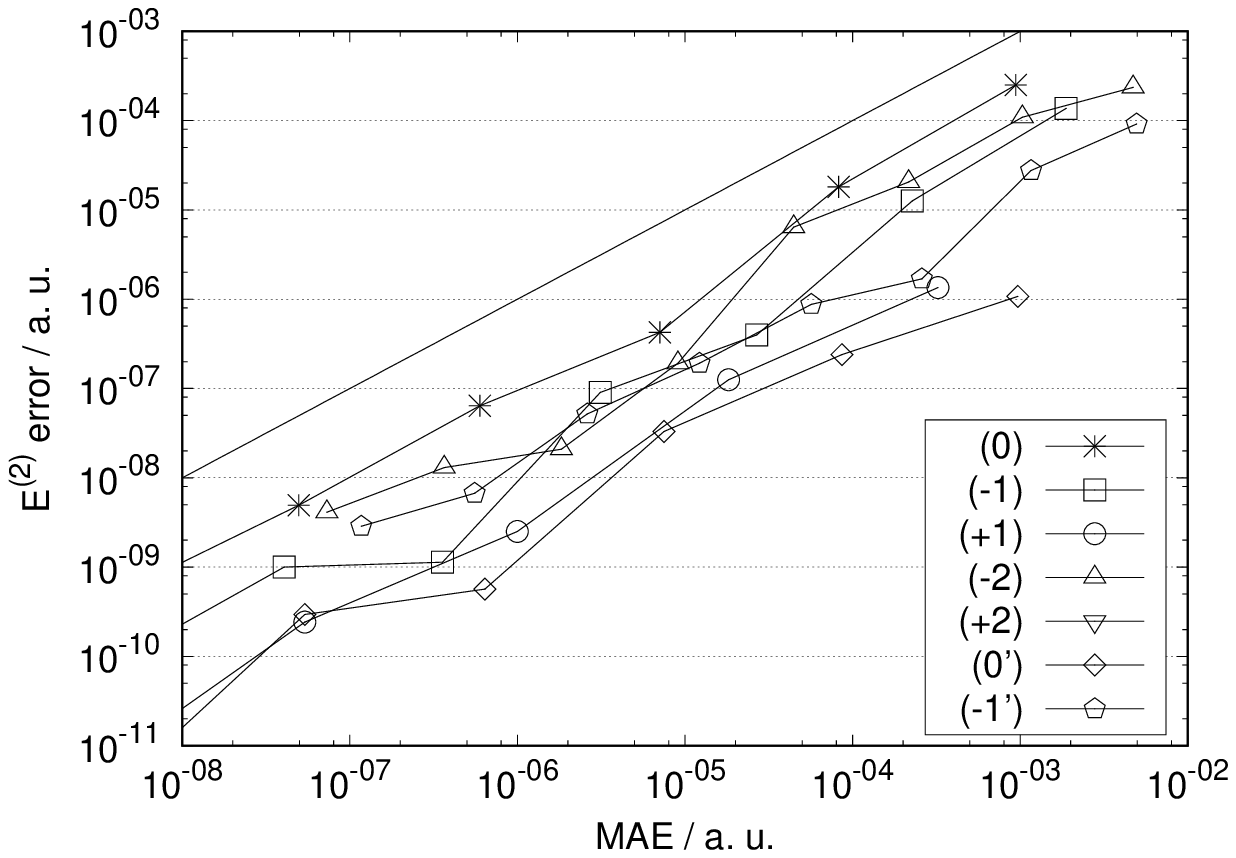}
 }
 \caption{
Numerical integration errors in Laplace-transformed AO-NEVPT2 energies
for Cl\textsubscript{2} (\mbox{r\tief{eq} = 1.9879~\AA}, CAS (14,8),
cc-pwCVTZ basis set\cite{}, 1s orbitals frozen)
with respect to the number of quadrature points \subref{fig1a}
and the maximum absolute error of Eq.\ \eqref{edf} in \subref{fig1b}.
}
 \label{fig1}
\end{figure}

\begin{figure}[htbp]
 \centering
 \includegraphics[width=0.95\textwidth]{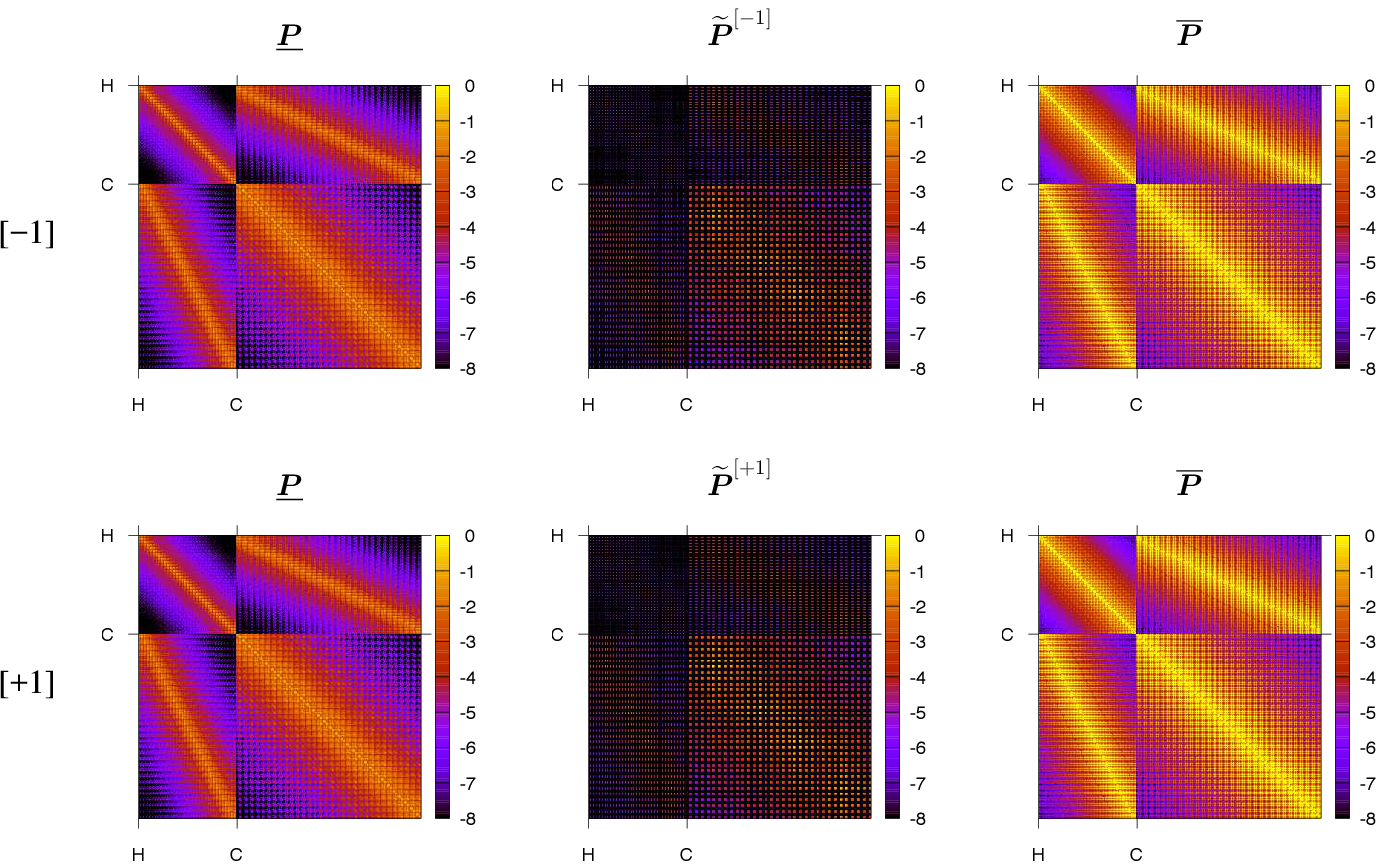}
 \caption{
Maximum norm for each shell pair of core $\boldsymbol{\po}$,
active $\boldsymbol{\pa}$, and virtual pseudo-density matrices $\boldsymbol{\pv}$
of the $\spB$ and $\spC$ contribution in the linear polyene chains
$\text{C}_{32}\text{H}_{34}$.
A CAS (8,8) space, SVP basis set,\cite{Schaefer1992} and a single quadrature point ($n_{\alpha}=1$) were used.
}
 \label{fig2}
\end{figure}

\begin{figure}[htbp]
 \centering
 \includegraphics[width=0.95\textwidth]{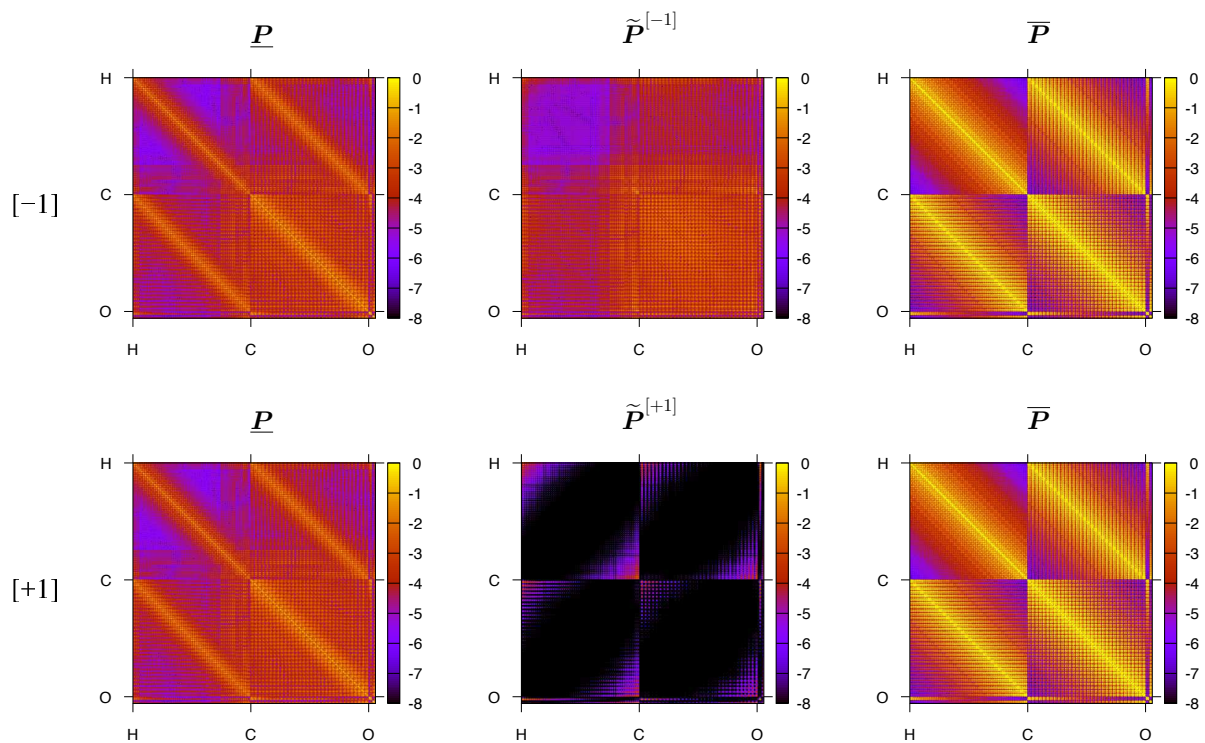}
 \caption{
Maximum norm for each shell pair of core $\boldsymbol{\po}$,
active $\boldsymbol{\pa}$, and virtual pseudo-density matrices $\boldsymbol{\pv}$
of the $\spB$ and $\spC$ contribution in the linear all-trans
polyethylene glycol
$\dot{\text{O}}-\text{C}_{32}\text{H}_{64}-\dot{\text{O}}$ biradical.
A CAS (10,6) space, SVP basis set,\cite{Schaefer1992} and a single quadrature point ($n_{\alpha}=1$) were used.
}
 \label{fig3}
\end{figure}

The presence of three different orbital spaces results in eight
different contributions to the IC first-order wavefunction contribution
with eight different OEDs.
Each of the 8~OEDs is approximated by its own
numerical quadrature of its Laplace transform,
which is determined by minimization the maximum absolute error (MAE)
of the error distribution function\cite{Takatsuka2008}
\begin{align}
\eta(\Delta) &= \frac{1}{\Delta} -
 \sum_{\alpha=1}^{n_{\alpha}} \omega_{{\alpha}}\, \exp(-\Delta\, t_{{\alpha}}) \label{edf}
\text{.}
\end{align}
This procedure is known as the minimax algorithm\cite{Takatsuka2008,Helmich-Paris2016d}
and can be performed in interval $[1,R]$ with
$R=\Delta_{\text{max}}/\Delta_{\text{min}}$ by scaling
the parameters $\{ \omega_{\alpha}, t_{\alpha} \}$.
This facilitates pre-tabulation of $\{ \omega_{\alpha}, t_{\alpha} \}$
along with their MAEs $\delta_{n_{\alpha},[1,R']}$ for some discrete $R'$.
These $2 n_{\alpha}+1$ parameters can be used as start values
in the iterative optimization procedure.\cite{Laplace-Repo}
We choose $n_{\alpha}$ such that
\begin{align} \label{lapcond}
& \delta_{n_{\alpha},[1,R']} \le T_{\text{lap}}
 \text{ with }
R\le R'
\end{align}
where $R'$ is the pre-tabulated range closest to $R$.
The accuracy of the numerical quadrature depends on a
single user-given threshold $T_{\text{lap}}$ that should correlate
with the resultant error in each correlation energy contribution.
Alternatively, one could simply set $n_{\alpha}$ to the same value
for each of the 8~OEDs based on experience.
However, by doing so, the errors in $E^{(2)}$ cannot be easily
controlled for each class.
This can be clearly seen from Fig.\ \ref{fig1a} where
the 8 different $E^{(2)}$ errors differ from each other
even by more than 4 orders of magnitude for some $n_{\alpha}$.
Conversely, the same $E^{(2)}$ errors converge much more uniformly
with respect to the MAE $\delta_{n_{\alpha},[1,R]}$,
as shown in Fig.\ \ref{fig1b}.

In the following, we investigate the $E^{(2)}$ errors when $n_{\alpha}$
is selected according to \eqref{lapcond} for the ground
state ${^1}\Sigma_g^{+}$ of the fluorine and chlorine dimer.
The results for different computational setups are compiled in Tab.\ \ref{tab4}.
For F\tief{2} and Cl\tief{2} (a multiple of) the equilibrium bond distances
$(n\times ) r_{eq}$ were taken from experimental data.\cite{nistdiatom}
For F the cc-pVTZ\cite{Dunning1989} and aug-cc-pVTZ\cite{Kendall1992}
basis sets were employed; for Cl we used the cc-pwCVTZ\cite{Woon1993}
basis set.
The Laplace accuracy threshold $T_{\text{lap}}$ was set to $10^{-7}$.
We note that for all calculations in Tab.\ \ref{tab4} the $\spG$
part of the $E^{(2)}$ is zero and is not shown.

For the first two $E^{(2)}$ calculation of F\tief{2} in Tab.\ \ref{tab4}
we chose a CAS (10,6) that includes all $2p$ orbitals and electrons of F.
The first calculation differs from the second only
in correlation of the 1s core orbitals.
The 8 OED ranges $R$ of the all-electron (AE) calculation varies roughly
by a factor of two only and similar $n_{\alpha}$,
i.\ e.\ 7 and 8, are selected by our procedure.
The absolute errors in $E^{(2)}$ are all below $10^{-9}$.
By freezing the $1s$ orbitals the OED ranges of those
classes that involve
core orbitals are decreased significantly while the
$\spD$ and $\spF$ contributions remain unchanged.
The latter require the least $n_{\alpha}$ (7) in the AE calculation
but the most in the frozen-core (FC) calculation.
For the FC calculation the OED ranges differ by almost 1~order
of magnitude. Nevertheless, all absolute errors in $E^{(2)}$
are smaller than $10^{-8}$.
Augmenting the cc-pVTZ basis by one diffuse basis function for
each angular momentum\cite{Kendall1992} lowers the eigenvalues
of the virtual-virtual block of the Fock matrix.
We can observe for the third calculation in Tab.\ \ref{tab4}
that the OED ranges of those classes that involve virtual
orbitals are increased approximately by a factor of 2
while those of the $\spE$ space are almost unaffected
when adding the diffuse ``aug-'' functions to the basis set.
The absolute errors in $E^{(2)}$ are still reliably
below $10^{-9}$.

For Cl atom in Cl\tief{2} we correlate the $2s2p3s3p$ orbitals
and keep the $1s$ frozen.
For the fourth calculation in Tab.\ \ref{tab4}
we put the all $3p$ orbitals and electrons of Cl
into the active space, that is CAS (10,6).
For this calculation the OED ranges become much larger than
those of the FC F\tief{2} calculations, which can
be attributed to the additional core electrons
and the larger basis set for Cl.
The largest OED range in the fourth calculation
is 114.76, while the largest in F\tief{2} calculation
with the aug-cc-pVTZ basis set is 24.32.
Though the OED ranges vary by more than 1~order of
magnitude for the Cl\tief{2} calculation,
all absolute errors in $E^{(2)}$ are less than $10^{-8}$
and $n_{\alpha}$ seems to be selected properly.
In the fifth calculation in Tab.\ \ref{tab4}
the $3s$ orbitals and electrons of the Cl atoms
are included in the valence space, i.\ e.\ CAS (14,8).
This lowers substantially the OED ranges of all those
classes that excite from core electrons.
The smallest ($\spE$) and the largest OED range ($\spF$) differ
by almost 2~orders of magnitude but $n_{\alpha}$
is selected properly as the absolute error in $E^{(2)}$
is always smaller $10^{-8}$.
The convergence of the absolute $E^{(2)}$ error
with respect to $n_{\alpha}$ and the MAE $\delta_{n_{\alpha},[1,R]}$
is shown in Fig.\ \ref{fig1} and has been discussed already
in the beginning of this section.

The previous example calculations of F\tief{2} and
Cl\tief{2} had clearly single-reference character
as deduced from their natural occupation numbers ($n_{\text{NO}} > 1.85$
or $n_{\text{NO}} < 0.15$) and could have been performed more easily with
closed-shell single-reference MP2 or coupled cluster.
Therefore, we performed the CAS (14,8) calculation
also on a stretched Cl\tief{2} with $3\times r_{eq}$ bond distance.
Now there are 2~open shells with $n_{\text{NO}} = 1.00$
and 6~closed shells with $n_{\text{NO}} = 2.00$.
As we investigate the ${^1}\Sigma^{+}_g$ state of Cl\tief{2},
at least 2~CSF are required to describe the system at least
qualitatively correct.
Thus, he have a MR case by definition.
The MR character does not seem to
change the OEDs much and we obtain absolute errors in $E^{(2)}$
that are all smaller than $10^{-8}$.
It is noteworthy that the $\spE$ contribution
to the correlation energy almost vanishes
since Cl\tief{2} can be considered as nearly
dissociated at $3\times r_{eq}$
and it is impossible to attach 2~electrons into a
CAS (5,3) or (7,4) valence space of a single Cl atom.

A further field of application of MRPT methods are excited states.
With our state-specific (SS) Laplace-transformed PC-NEVPT2 method
we investigate the OED ranges and absolute errors
in $E^{(2)}$ of the ground state 1\,\hoch{1}A\tief{1}
and two lowest vertically excited singlet states
1\,\hoch{1}A\tief{2} ($n_y\to \pi^*$) and 1\,{\hoch{1}}B\tief{2} ($n_y\to 3s$)
in formaldehyde,
which are shown in Tab.\ \ref{tab5}.
Those and other valence and Rydberg excited states have been investigated rigorously
by Merch{\'a}n and Roos\cite{Merchan1995} with SS-CASPT2 and by Angeli et al.\
with SS-NEVPT2\cite{Angeli2004b}.
The geometry and the ANO(S)-Ryd(S) basis set were taken from their works.
To describe also Rydberg states, a contracted s, p, and d function,
denoted by Ryd(S), were located altogether in the geometric center of H\tief{2}CO.
In contrast to Refs.\ \onlinecite{Merchan1995}
and \onlinecite{Angeli2004b} we chose a CAS (4,4) that includes the $(1b_1, 2b_2, 6a_1, 2b_1)$
orbitals to optimize the orbitals and CI coefficients of the
1\,\hoch{1}A\tief{1}, 1\,\hoch{1}A\tief{2}, and 1\,{\hoch{1}}B\tief{2} states
simultaneously in the SA-CASSCF procedure because,
currently, we cannot exploit PG symmetry for the NEVPT2 calculations
with our pilot implementation.
Nevertheless,
the excitation energies to the
1\,\hoch{1}A\tief{2} (4.10~eV) and 1\,{\hoch{1}}B\tief{2} state (7.18~eV)
differ only by -0.07 and +0.10~eV, respectively,
though an active space was employed that differs from
the ones of Ref.\ \onlinecite{Angeli2004b}.

As observed already for the halogen dimers,
those classes that attach electrons into the active space have
the smallest OED ranges and demand the least $n_{\alpha}$
to reach a desired accuracy for all 3~states in Tab.\ \ref{tab5};
while active space electron ionization classes have
larger OED ranges and need the most $n_{\alpha}$.
The combined single electron ionization and excitation
class $\spF$ has the largest OED range for each of the 3~states but
does not contribute with more than 7~\% to the total $E^{(2)}$ energy.
For the Rydberg excited state 1\,{\hoch{1}}B\tief{2}
the OED range of the $\spF$ class is only a factor of two larger than for
both the ground
state 1\,\hoch{1}A\tief{1} and first valence excited state 1\,\hoch{1}A\tief{2}.
Furthermore, only 1~additional quadrature point
is required for $\spF$ in 1\,{\hoch{1}}B\tief{2} to reach the requested target accuracy.
Therefore, OED ranges of $E^{(2)}$ contributions to
Rydberg excited states are larger than for valence excited states;
but even if diffuse basis functions are included in the basis set,
the additional number of quadrature points is still rather decent.
In fact, it is the less effective screening of shell contributions
in the AO basis rather than the additional number of quadrature points
that makes an AO-based formulation less powerful for
basis sets with diffuse basis functions in a production-level implementation.
This we will study in future work.


\section{Conclusions}
In the present article we showed how to formulate the partially 
contracted n-electron valence
second order perturbation theory (PC-NEVPT2) energies in the atomic and active
molecular orbital basis by employing the Laplace transformation
of orbital-energy denominators (OED).
As the number of active orbitals is comparatively small and 
AO basis functions are inherently localized, our formulation is 
particularly suited for a linearly-scaling NEVPT2 implementation.
Some of the NEVPT2 energy contributions can be formulated completely
in the AO basis as single-reference second-order M{\o}ller--Plesset perturbation
theory and benefit from sparse active-pseudo 
density matrices --- particularly if the active molecular orbitals are localized
only in parts of a molecule.
Furthermore, we could show that finding optimal parameters of the 
numerical Laplace transformation
is particularly challenging for multi-reference perturbation theories 
as the fit range may vary among the 8~different OEDs
by many orders of magnitude.
Selecting the number of quadrature points for each OED
separately according to an accuracy-based criterion allows us
to control the errors in the NEVPT2 energies reliably.
Currently, we work on efficient low-order scaling implementations
of NEVPT2 and relativistic two-component MP2.
The extension of our formalism to quasi-degenerate NEVPT2\cite{Angeli2004}
or the uncontracted time-dependent NEVPT2\cite{Sokolov2016} will
be presented in the future.

\section{Acknowledgments}
This work was supported by the
the German research foundation DFG
through a research fellowship (Grant No.\ HE 7427/1-1)
and the Dutch science foundation NWO
through a VENI fellowship (Grant No.\ 722.016.011).
B.\ H.-P.\ would like to thank Markus Reiher for granting access to
the local computing facilities at ETH Z{\"u}rich and the NWO
for computer time at the Dutch national supercomputers.
The authors also thank Hans J{\o}rgen {\relax Aa}.\ Jensen
for sharing his experiences in multi-reference electronic structure methods.

\bibliography{paper}

\begin{table}[h]
\caption{
Explicit expressions for the Laplace-transformed PC-NEVPT2
energies and their intermediates in the atomic and
active molecular orbital basis.
}
\label{tab1}
\begin{ruledtabular}
\begin{tabular}{l}
\[ E^{\spA} = - \sum_{\ga}\, \sum_{\gm\gn\gk\gl} (\ogm \vgn|\gk\gl)^{(\ga)}
  \Big[ 2 (\gm\gn|\ogk\vgl)^{(\ga)} - (\gm\vgl|\ogk\gn)^{(\ga)} \Big] \] \\[1.0em]
\[ E^{\spB} = - \sum_{\ga} \sum_{\gm\gn\gk\gl}  (\ogm\vgn|\gk\gl)^{(\ga)}
 \Big[ 2 (\gm \gn|\agk^{\spB} \vgl)^{(\ga)} - (\gm\vgl|\agk^{\spB}\gn)^{(\ga)} \Big] \]  \\[1.0em]
\[ E^{\spC} = - \sum_{\ga} \sum_{\gm\gn\gk\gl} (\ogm\vgn|\gk\gl)^{(\ga)}
 \Big[ 2 (\gm \gn|\ogk \agl^{[+1]})^{(\ga)} - (\gm \agl^{[+1]}|\ogk\gn)^{(\ga)} \Big] \] \\[1.0em]
\[ E^{\spD} = -\frac{1}{2} \sum_{\ga} \sum_{\substack{\gn\gl \\ tut'u'}} (t\gn|u\gl)\,
 \mathcal{K}^{(\ga),\spD}_{tu,t'u'}\,   (t'\vgn|u'\vgl)^{(\ga)} \] \\[1.0em]
\[ E^{\spE} = -\frac{1}{2} \sum_{\ga} \sum_{\substack{\gm\gk \\ tut'u'}} (\gm t|\gk u)\,
 \mathcal{K}^{(\ga),\spE}_{tu,t'u'}\,  (\ogm t'|\ogk u')^{(\ga)} \] \\[1.0em]
\[ E^{\spF} = -\sum_{\ga} \sum_{\substack{ \gm \\ tuvt'u'v'}} \widetilde{(\gm u|tv)}^{\spF}\,
 \mathcal{K}^{(\ga),\spF}_{tuv,t'u'v'} \widetilde{({\vgm} u'|t'v')}^{(\ga),\spF} \] \\[1.0em]
\[ E^{\spG} = -\sum_{\ga} \sum_{\substack{\gm \\ tuvt'u'v'}} \widetilde{(\gm u|tv)}^{\spG}\,
 \mathcal{K}^{(\ga),\spG}_{tuv,t'u'v'}\, \widetilde{(\ogm u'|t'v')}^{(\ga),\spG} \] \\[1.0em]
\[ E^{\spH} = -\sum_{\ga} \sum_{\substack{ \gm\gn \\ tut'u'}} \Big[ \widetilde{(\gm\gn|tu)}^{\spH}\,
 \mathcal{K}^{(\ga),\spH,AA}_{tu,t'u'}\, \widetilde{(\ogm\vgn|t'u')}^{(\ga),\spH} \] \\[1.0em]
\[ \phantom{E^{\spH} = -\sum_{\ga} \sum_{\substack{\gm\gn \\ tut'u'}}} + 2 \widetilde{(\gm\gn|tu)}^{\spH}\,
 \mathcal{K}^{(\ga),\spH,AB}_{tu,t'u'}\, (\ogm u'|t'\vgn)^{(\ga)} \] \\[1.0em]
\[ \phantom{E^{\spH} = -\sum_{\ga} \sum_{\substack{\gm\gn \\ tut'u'}}} +  (\gm u|t\gn)\,
 \mathcal{K}^{(\ga),\spH,BB}_{tu,t'u'}\, (\ogm u'|t'\vgn)^{(\ga)} \Big] \] \\[1.0em]
\hline
\[ (\gm \gn|\ogk \agl^{[+1]})^{(\ga)} =
 \sum_{\gk'} \po_{\gk\gk'}^{(\ga)} \sum_{\gl'} \pa_{\gl\gl'}^{(\ga),\spC}\, (\gm \gn| \gk'\gl') \] \\[1.0em]
\[\widetilde{(\gm\gn|tu)}^{\spH} =
 (\gm\gn|tu) + f_{\gm\gn}^{\text{I}} \sum_{vw} \1rdm_{vw} [(\mathbf{M}^{\spH})^{-1}]_{tu,vw} \] \\[1.0em]
\[\widetilde{(\gm t|uv)}^{\spF} =
 (\gm t|uv) + \sum_w f'^{I}_{\gm w} \sum_{tuv} (\2rdm_{vu,tw} + \gd_{tu}\, \1rdm_{vw} )\,
   \sum_{\gt} c_{tuv,\gt}^{\spF}\, c_{tuv,\gt}^{\spF} \] \\[1.0em]
\[ \widetilde{(\gm t|uv)}^{\spG}
= (\gm t|uv) + \sum_w f^{I}_{\gm w} \sum_{tuv} (2\gd_{uw} \1rdm_{vt} - \gd_{tw} \1rdm_{vu} - \2rdm_{vw,tu} )\,
  \sum_{\gt} c_{tuv,\gt}^{\spG}\, c_{tuv,\gt}^{\spG} \] \\[1.0em]
\[ f'^{\text{I}}_{\gm t} = f^{\text{I}}_{\gm t} - (\gm u|u t) \] \\[1.0em]
\[ \pa_{\gm\gm'}^{(\ga),\spC} =  \sum_{t} C_{\gm t}\, \sum_{t'} \mathcal{K}^{(\ga),\spC}_{tt'}\, C_{\gm'u'} \]
\end{tabular}
\end{ruledtabular}
\end{table}

\begin{table}[h]
\caption{
 Metric matrices for each class (top) formulated
 in terms of singlet reduced density matrices (bottom)
}
\label{tab2}
\begin{ruledtabular}
\begin{tabular}{l}
\[ M^{\spB}_{tt'} = \1rdm_{tt'} \] \\[0.5em]
\[ M^{\spC}_{tt'} = 2\gd_{tt'} - \1rdm_{t't} \] \\[0.5em]
\[ M^{\spD}_{tu,t'u'} = \2rdm_{tu,t'u'} \] \\[0.5em]
\[ M^{\spE}_{tu,t'u'} = \2rdm_{tu,t'u'} + 4\, \gd_{tt'}\,\gd_{uu'}  - 2\, \gd_{tu'}\,\gd_{ut'} \] \\[0.5em]
\[ \phantom{M^{\spE}_{tu,t'u'} =} - 2\, \gd_{tt'}\, \1rdm_{uu'}  + \gd_{tu'}\, \1rdm_{ut'} - 2\, \gd_{uu'}\, \1rdm_{tt'}  + \gd_{ut'}\, \1rdm_{tu'} \] \\[0.5em]
\[ M^{\spF}_{tuv,t'u'v'} = \gd_{tu}\, \gd_{t'u'}\, \1rdm_{vv'} + \gd_{t'u'}\, \2rdm_{vu,tv'} + \gd_{tu}\, \2rdm_{vt',u'v'} \] \\[0.5em]
\[ \phantom{M^{\spF}_{tuv,t'u'v'} =} + \gd_{tt'}\, \2rdm_{vu,v'u'} + \3rdm_{vut',tu'v'} \] \\[0.5em]
\[ M^{\spG}_{tuv,t'u'v'} = 2\,\gd_{tt'}\,\gd_{uu'}\, \1rdm_{vv'} + 2\, \gd_{uu'}\, \2rdm_{vt',tv'} - M^{\spF}_{tu'v,t'uv'} \] \\[0.5em]
\[ \mathbf{M}^{\spH} = \left[
 \begin{array}{cc}
 2 \mathbf{M}^{AA} & -\mathbf{M}^{AB} \\
 - \mathbf{M}^{BA} &  \mathbf{M}^{BB}
 \end{array}
\right]
 \] \\[0.5em]
\[ \mathbf{M}^{AA} = \mathbf{M}^{AB} = \mathbf{M}^{BA} =
  M^{\spH}_{tu,t'u'} = \gd_{uu'}\, \1rdm_{tt'} + \2rdm_{tu',ut'} \] \\[0.5em]
\[ \mathbf{M}^{BB} = M'^{\spH}_{tu,t'u'} = 2\, \gd_{uu'}\, \1rdm_{tt'} - \2rdm_{u't,ut'} \] \\[0.5em]
\hline
\[ \1rdm_{tt'} = \sum_{\gs} \bra{0} \acr_{t\gs} \aan_{t'\gs} \ket{0} \] \\[0.5em]
\[ \2rdm_{tu,t'u'} = \sum_{\gs\gs'} \bra{0} \acr_{t\gs} \acr_{u\gs'} \aan_{u'\gs'} \aan_{t'\gs} \ket{0} \] \\[0.5em]
\[ \3rdm_{tuv,t'u'v'} = \sum_{\gs\gs'\gs''} \bra{0} \acr_{t\gs} \acr_{u\gs'} \acr_{v\gs''} \aan_{v'\gs''} \aan_{u'\gs'} \aan_{t'\gs} \ket{0} \] \\[0.5em]
\[ \4rdm_{tuvw,t'u'v'w'} = \sum_{\gs\gs'\gs''\gs'''}
  \bra{0} \acr_{t\gs} \acr_{u\gs'} \acr_{v\gs''} \acr_{w\gs'''}
          \aan_{w'\gs'''} \aan_{v'\gs''} \aan_{u'\gs'} \aan_{t'\gs} \ket{0} \]
\end{tabular}
\end{ruledtabular}
\end{table}

\begin{table}
\caption{
Definition of the internally contracted configurations (ICC) for the 8 excitation classes.
For CASPT2 and PC-NEVPT2 the projection manifold is built from a bi-orthogonal ICC basis
while for SC-NEVPT2 a special basis is chosen with contraction over all active orbitals.
}
\label{tab3}
\begin{ruledtabular}
\begin{tabular}{llll}
\multicolumn{1}{c}{label} &
\multicolumn{1}{c}{$\ket{\text{ICC}}$} &
\multicolumn{1}{c}{$\bra{\text{ICC}}$} &
\multicolumn{1}{c}{SC $\bra{\text{ICC}}$} \\
 \hline
 $ \spA $ & $ \ket{\Phi_{ij}^{ab}} = \hat{E}_{ai} \hat{E}_{bj} \ket{0}$  &
 $\bra{\widetilde{\Phi_{ij}^{ab}}} =
  \frac{1}{6} \bra{0} \Big( 2\, \hat{E}_{jb} \hat{E}_{ia} +
                                \hat{E}_{ib} \hat{E}_{ja} \Big)$ & \\
 $ \spB $ & $ \ket{\Phi_{it}^{ab}} = \hat{E}_{ai} \hat{E}_{bt} \ket{0} $  &
 $\bra{\widetilde{\Phi_{it}^{ab}}} =
  \frac{1}{3} \bra{0} \Big( 2\, \hat{E}_{tb} \hat{E}_{ia} +
                                \hat{E}_{ib} \hat{E}_{ta} \Big)$ &
 $\bra{\widetilde{\Phi_{i}^{ab}}} =
  \frac{1}{3} \sum_t g_{tbia} \bra{0} \Big( 2\, \hat{E}_{tb} \hat{E}_{ia} +
                                                \hat{E}_{ib} \hat{E}_{ta} \Big)$
\\
 $ \spC $ & $ \ket{\Phi_{ij}^{at}} = \hat{E}_{ai} \hat{E}_{tj} \ket{0} $  &
 $\bra{\widetilde{\Phi_{ij}^{at}}} =
  \frac{1}{3} \bra{0} \Big( 2\, \hat{E}_{jt} \hat{E}_{ia} +
                                \hat{E}_{it} \hat{E}_{ja} \Big)$ &
 $\bra{\widetilde{\Phi_{ij}^{a}}} =
  \frac{1}{3} \sum_t  g_{jtia} \bra{0} \Big( 2\, \hat{E}_{jt} \hat{E}_{ia} +
                                                 \hat{E}_{it} \hat{E}_{ja} \Big)$ \\
  $ \spD $ & $\ket{\Phi_{tu}^{ab}} = \hat{E}_{at} \hat{E}_{bu} \ket{0}$  &
 $\bra{\widetilde{\Phi_{tu}^{ab}}} = \bra{0} \hat{E}_{ub} \hat{E}_{ta}$  &
 $\bra{\widetilde{\Phi^{ab}}}      = \sum_{tu} g_{ubta} \bra{0} \hat{E}_{ub} \hat{E}_{ta}$  \\
  $ \spE $ & $\ket{\Phi_{ij}^{tu}} = \hat{E}_{ti} \hat{E}_{uj} \ket{0}$  &
 $\bra{\widetilde{\Phi_{ij}^{tu}}} = \bra{0} \hat{E}_{ju} \hat{E}_{ti}$  &
 $\bra{\widetilde{\Phi_{ij}}}      = \sum_{tu} g_{juti} \bra{0} \hat{E}_{ju} \hat{E}_{ti}$  \\
  $ \spF $ & $\ket{\Phi_{tu}^{av}} = \hat{E}_{at} \hat{E}_{vu} \ket{0}$  &
 $\bra{\widetilde{\Phi_{tu}^{av}}} = \bra{0} \hat{E}_{uv} \hat{E}_{ta}$  &
 $\bra{\widetilde{\Phi^{a}}} = \sum_{tuv} g_{uvta} \bra{0} \hat{E}_{uv} \hat{E}_{ta}
                                 + \sum_t f'^{\text{I}}_{ta} \bra{0} \hat{E}_{ta} $\\
  $ \spG $ & $\ket{\Phi_{it}^{uv}} = \hat{E}_{ui} \hat{E}_{vt} \ket{0}$  &
 $\bra{\widetilde{\Phi_{it}^{uv}}} = \bra{0} \hat{E}_{tv} \hat{E}_{iu}$  &
 $\bra{\widetilde{\Phi_{i}}}       = \sum_{tuv} g_{tviu} \bra{0} \hat{E}_{tv} \hat{E}_{iu}
                                   + \sum_t  f^{\text{I}}_{it} \bra{0} \hat{E}_{it}$\\
 $ \spH $ & $ \ket{\Phi_{it}^{au}} = \hat{E}_{ai} \hat{E}_{ut} \ket{0}$, &
 $\bra{\widetilde{\Phi_{it}^{au}}} = \bra{0} \hat{E}_{tu} \hat{E}_{ia}$, &
 $\bra{\widetilde{\Phi_{i}^{a}}} =  \sum_{tu} \Big( g_{tuia} \bra{0} \hat{E}_{tu} \hat{E}_{ia}
  +  g_{taiu} \hat{E}_{ta} \hat{E}_{iu} \Big)$ \\
          & $ \ket{\Phi_{it}^{ua}} = \hat{E}_{ui} \hat{E}_{at} \ket{0}$  &
 $\bra{\widetilde{\Phi_{it}^{ua}}} = \bra{0} \hat{E}_{ta} \hat{E}_{iu}$  &
 $\phantom{\bra{\widetilde{\Phi_{i}^{a}}} =} + f^{\text{I}}_{ia} \bra{0} \hat{E}_{ia}$
\end{tabular}
\end{ruledtabular}
\end{table}

\begin{table}
\caption{
Errors ($\Delta E$) of the numerical quadrature
and reference ($E_{\text{ref}}$)
for each PC-NEVPT2 energy contribution of
Cl\tief{2} and F\tief{2} in the ground state ${^1}\Sigma_g^+$.
The target accuracy for the determination of the
number of quadrature points $n_{\alpha}$
was set to $T_{\text{lap}} = 10^{-7}$.
}
\label{tab4}
\begin{ruledtabular}
\begin{tabular}{crrrr}
\multicolumn{1}{c}{class} &
\multicolumn{1}{c}{$E_{\text{ref}}$ / a.\ u.} &
\multicolumn{1}{c}{$\Delta E$ / a.\ u.} &
\multicolumn{1}{c}{$R$} &
\multicolumn{1}{c}{$n_{\alpha}$} \\[0.5em]
 \hline
\multicolumn{5}{c}{F\tief{2} (r\tief{eq}), cc-pVTZ, CAS (10,6),	AE} \\[0.5em]
%
%
$\spA$ &-0.030025488553 & $ 3.17 \times 10^{-10}$ & 16.37 & 8 \\
$\spB$ &-0.111943993847 & $ 7.13 \times 10^{-10}$ & 13.29 & 8 \\
$\spC$ &-0.002340179047 & $ 3.00 \times 10^{-11}$ & 16.13 & 8 \\
$\spD$ &-0.212848549931 & $-8.08 \times 10^{-10}$ & 7.47  & 7 \\
$\spE$ &-0.001439019119 & $ 1.00 \times 10^{-11}$ & 12.99 & 8 \\
$\spH$ &-0.047509476974 & $-6.97 \times 10^{-10}$ & 16.90 & 8 \\
$\spF$ &-0.043202501004 & $ 6.10 \times 10^{-10}$ & 9.44  & 7 \\
sum    &-0.449309208510 & $ 1.75 \times 10^{-10}$ & &         \\[0.75em]
\multicolumn{5}{c}{F\tief{2} (r\tief{eq}), cc-pVTZ, CAS (10,6),	1s frozen} \\[0.5em]
$\spA$ &-0.018579052395 & $-1.00 \times 10^{-11}$ & 5.53 & 6 \\
$\spB$ &-0.100501783485 & $-3.87 \times 10^{-09}$ & 6.65 & 6 \\
$\spC$ &-0.001394951164 & $ 4.80 \times 10^{-11}$ & 3.78 & 5 \\
$\spD$ &-0.212848549931 & $-8.08 \times 10^{-10}$ & 7.47 & 7 \\
$\spE$ &-0.001368251581 & $<10^{-12}$             & 1.12 & 3 \\
$\spH$ &-0.046541048840 & $-2.84 \times 10^{-10}$ & 6.05 & 6 \\
$\spF$ &-0.043202501004 & $ 6.10 \times 10^{-10}$ & 9.44 & 7 \\
sum    &-0.424436138435 & $-4.31 \times 10^{-09}$ &      &   \\[0.75em]
\multicolumn{5}{c}{F\tief{2} (r\tief{eq}), aug-cc-pVTZ, CAS (10,6),	1s frozen} \\[0.5em]
$\spA$ &-0.018942769474	& $4.80 \times 10^{-11}$ & 10.65 & 8 \\
$\spB$ &-0.104717284874	& $9.60 \times 10^{-10}$ & 14.15 & 8 \\
$\spC$ &-0.001432889156	& $2.00 \times 10^{-12}$ & 5.93  & 6 \\
$\spD$ &-0.218661812847	& $5.58 \times 10^{-09}$ & 16.88 & 8 \\
$\spE$ &-0.001372612217	& $<10^{-12}$            & 1.12  & 3 \\
$\spH$ &-0.046967568913	& $5.60 \times 10^{-11}$ & 11.39 & 8 \\
$\spF$ &-0.044436625063	& $3.16 \times 10^{-10}$ & 24.32 & 9 \\
sum    &-0.436531562544	& $6.96 \times 10^{-09}$ & 	     &   \\[0.75em]
\multicolumn{5}{c}{Cl\tief{2} (r\tief{eq}), cc-pwCVTZ, CAS (10,6),	1s frozen} \\[0.5em]
$\spA$ &-0.471288060730 & $ 5.03 \times 10^{-09}$ & 72.20 & 10 \\
$\spB$ &-0.156827410078 & $-5.59 \times 10^{-09}$ & 86.56 & 10 \\
$\spC$ &-0.015084015071 & $ 1.26 \times 10^{-09}$ & 43.77 & 9  \\
$\spD$ &-0.178487055950 & $-8.42 \times 10^{-09}$ & 93.65 & 10 \\
$\spE$ &-0.001461454900 & $ 7.00 \times 10^{-12}$ &  7.93 & 7  \\
$\spH$ &-0.042764807405 & $ 4.76 \times 10^{-10}$ & 72.86 & 10 \\
$\spF$ &-0.028430997436 & $-2.44 \times 10^{-10}$ &114.76 & 11 \\
sum    &-0.894343801571 & $-7.47 \times 10^{-09}$ & 	  &    \\[0.75em]
\multicolumn{5}{c}{Cl\tief{2} (r\tief{eq}), cc-pwCVTZ, CAS (14,8),	1s frozen} \\[0.5em]
$\spA$ &-0.423057258482	& $ 4.96 \times 10^{-10}$ & 11.86 & 8   \\
$\spB$ &-0.108905246973	& $ 1.01 \times 10^{-10}$ & 20.62 & 9   \\
$\spC$ &-0.009006319960	& $ 2.42 \times 10^{-10}$ &  6.71 & 6   \\
$\spD$ &-0.261212365274	& $-4.11 \times 10^{-09}$ & 94.37 & 10  \\
$\spE$ &-0.000139138179	& $ 1.00 \times 10^{-12}$ &  1.30 & 3   \\
$\spH$ &-0.007634923746	& $ 5.00 \times 10^{-12}$ & 12.10 & 8   \\
$\spF$ &-0.067654717180	& $-1.35 \times 10^{-10}$ &117.33 & 11  \\
sum    &-0.877609969794	& $-3.40 \times 10^{-09}$ &       &     \\[0.75em]
\multicolumn{5}{c}{Cl\tief{2} (3$\times$r\tief{eq}), cc-pwCVTZ, CAS (14,8),	1s frozen} \\[0.5em]
$\spA$ &-0.424002208658	& $ 4.54 \times 10^{-10}$ &11.64 & 8    \\
$\spB$ &-0.113366194595	& $-7.23 \times 10^{-10}$ &19.93 & 8    \\
$\spC$ &-0.005732731250	& $ 1.91 \times 10^{-10}$ &6.66  & 6    \\
$\spD$ &-0.250053179564	& $-7.97 \times 10^{-09}$ &90.82 & 10   \\
$\spE$ &-0.000000000704	& $<10^{-12}$             &1.31  & 3    \\
$\spH$ &-0.005920160948	& $-7.00 \times 10^{-12}$ &11.83 & 8    \\
$\spF$ &-0.049804208204	& $-3.75 \times 10^{-09}$ &97.24 & 10   \\
sum    &-0.848878684178	& $-1.18 \times 10^{-08}$ &	 &      
\end{tabular}
\end{ruledtabular}
\end{table}

\begin{table}
\caption{
Errors ($\Delta E$) of the numerical quadrature
and reference ($E_{\text{ref}}$)
for each PC-NEVPT2 energy contribution to the three
lowest singlet states of formaldehyde.
The natural occupation numbers
in the CAS (4,4) are provide as $(1b_1, 2b_2, 6a_1, 2b_1)$.
The target accuracy for the determination of the
number of quadrature points $n_{\alpha}$
was set to $T_{\text{lap}} = 10^{-7}$.
The ANO(S)-Ryd(S) basis set of Ref.\ \onlinecite{Angeli2004b}
was employed.
The $1s$ orbitals of the C and O atom were kept frozen
in the PT calculation.
}
\label{tab5}
\begin{ruledtabular}
\begin{tabular}{crrrr}
\multicolumn{1}{c}{class} &
\multicolumn{1}{c}{$E_{\text{ref}}$ / a.\ u.} &
\multicolumn{1}{c}{$\Delta E$ / a.\ u.} &
\multicolumn{1}{c}{$R$} &
\multicolumn{1}{c}{$n_{\alpha}$} \\[0.5em]
 \hline
\multicolumn{5}{c}{1{\hoch{1}}A\tief{1}, (1.90,2.00,0.00,0.10)} \\[0.5em]
$\spA$ &-0.100574701804	& $-9.00 \times 10^{-12}$ & 10.33 &8   \\
$\spB$ &-0.101361528657	& $-7.00 \times 10^{-12}$ & 11.86 &8   \\
$\spC$ &-0.012154365169	& $ 3.10 \times 10^{-11}$ & 6.51  &6   \\
$\spD$ &-0.035050155679	& $-4.55 \times 10^{-10}$ & 9.80  &7   \\
$\spE$ &-0.001562160519	& $ 2.00 \times 10^{-12}$ & 2.39  &5   \\
$\spH$ &-0.045498719711	& $-7.10 \times 10^{-11}$ & 12.00 &8   \\
$\spF$ &-0.021615019596	& $ 3.90 \times 10^{-11}$ & 20.34 &9   \\
$\spG$ &-0.000839589656	& $ 4.00 \times 10^{-12}$ & 4.24  &6   \\
sum    &-0.318656240791	& $-4.66 \times 10^{-10}$ & &          \\[0.75em]
\multicolumn{5}{c}{1\,{\hoch{1}}A\tief{2} (n\tief{y} $\to$ $\pi^*$), (2.00,1.00,0.00,1.00)} \\[0.5em]
$\spA$ &-0.100706612983	& $-5.00\times 10^{-12}$ & 10.48 & 8 \\
$\spB$ &-0.101295292993	& $-8.40\times 10^{-11}$ & 14.04 & 8 \\
$\spC$ &-0.019711077239	& $-8.20\times 10^{-11}$ & 6.44	 & 6 \\
$\spD$ &-0.030702543569	& $ 3.00\times 10^{-12}$ & 10.37 & 8 \\
$\spE$ &-0.000285995858	& $<10^{-12}$            & 2.40	 & 5 \\
$\spH$ &-0.046347363745	& $-1.24\times 10^{-10}$ & 16.16 & 8 \\
$\spF$ &-0.009305654325	& $-4.70\times 10^{-11}$ & 29.98 & 9 \\
$\spG$ &-0.003273183806	& $-6.90\times 10^{-11}$ & 5.19	 & 6 \\
tot    &-0.311627724518	& $-4.08\times 10^{-10}$ & 	 &   \\[0.75em]
\multicolumn{5}{c}{1\,{\hoch{1}}B\tief{2} (n\tief{y} $\to$ 3s), (1.98,1.00,1.00,0.02)} \\[0.5em]
$\spA$ &-0.100183219144	& $ 3.10 \times 10^{-11}$ & 8.11  & 7 \\
$\spB$ &-0.079052324949	& $-1.07 \times 10^{-10}$ & 13.61 & 8 \\
$\spC$ &-0.021486582338	& $-2.70 \times 10^{-11}$ & 5.80  & 6 \\
$\spD$ &-0.021022831398	& $-6.20 \times 10^{-11}$ & 11.90 & 8 \\
$\spE$ &-0.001895119491	& $<10^{-12}$             & 2.39  & 5 \\
$\spH$ &-0.064191826500	& $-1.25 \times 10^{-10}$ & 12.50 & 8 \\
$\spF$ &-0.020109659755	& $-7.89 \times 10^{-10}$ & 57.01 & 10\\
$\spG$ &-0.007188062195	& $-1.90 \times 10^{-11}$ & 4.71  & 6 \\
tot    &-0.315129625770	& $-1.10 \times 10^{-09}$ & 	  &   
\end{tabular}
\end{ruledtabular}
\end{table}

\end{document}